\documentclass{article}
\usepackage{bm,latexsym,amsmath,amssymb,amsfonts,fancyhdr,color,graphicx,multirow,slashed,cite}
\usepackage[a4paper,bottom=3cm,top=2.5cm,head=0mm,width=17cm,dvipdfm]{geometry}
\usepackage[usenames,dvipsnames,svgnames,table]{xcolor}
\usepackage[colorlinks=true,
            linkcolor=blue,
            urlcolor=blue,
            citecolor=green,          
						bookmarks=true,
						bookmarksnumbered=true,
						breaklinks=true,
						pdfpagemode=Fullscreen,
						pdfstartview=FitBH]{hyperref}
\usepackage[dotinlabels]{titletoc}
\usepackage{titlesec}
\usepackage{authblk,ulem}

\numberwithin{equation}{section}
\allowdisplaybreaks[4]

\titlelabel{\thetitle.\quad \hspace{-0.8em}}
\titlecontents{section}
              [1.5em]
              {\vspace{4mm} \large \bf}
              {\contentslabel{1em}}
              {\hspace*{-1em}}
              {\titlerule*[.5pc]{.}\contentspage}
\titlecontents{subsection}
              [3.5em]
              {\vspace{2mm}}
              {\contentslabel{1.8em}}
              {\hspace*{.3em}}
              {\titlerule*[.5pc]{.}\contentspage}
\titlecontents{subsubsection}
              [5.5em]
              {\vspace{2mm}}
              {\contentslabel{2.5em}}
              {\hspace*{.3em}}
              {\titlerule*[.5pc]{.}\contentspage}

\newcommand{\titledef}{New Physics Off the $Z$-Pole: $e^+ e^- \rightarrow f \bar f$ at Future Lepton Colliders\\ } % Insert Title here!!!
\hypersetup{ pdfauthor = {Shao-Feng Ge},
	     pdftitle = {\titledef}, % Insert title here!!!
	     pdfsubject = {}, % Insert subject here!!!
             pdfkeywords = {}, % Insert keywords here!!!
	     pdfcreator = {LaTeX with hyperref package},
	     pdfproducer = {dvips + ps2pdf} }

\definecolor{gesfblack}{rgb}{0,0,0}

\definecolor{gesfblue}{rgb}{0.08,0.42,0.76}

\definecolor{gesfgreen}{rgb}{0,1,0}

\definecolor{gesfgrey}{rgb}{0.5,0.5,0.5}

\definecolor{gesflanse}{rgb}{0.00,0.50,0.50}

\definecolor{gesfpurple}{rgb}{0.47,0.19,0.42}

\definecolor{gesfred}{rgb}{1,0,0}

\definecolor{gesfwhite}{rgb}{1,1,1}

\definecolor{gesfyellow}{rgb}{0.7,0.4,0.3}

\newcommand{\gsec}[1]{{\hypersetup{linkcolor=red}Sec.\,\ref{#1}\hypersetup{linkcolor=blue}}}

\newcommand{\geqn}[1]{\hypersetup{linkcolor=blue}Eq.\,(\ref{#1})\hypersetup{linkcolor=blue}}
\newcommand{\gfig}[1]{{\hypersetup{linkcolor=violet}Fig.\,\ref{#1}\hypersetup{linkcolor=blue}}}
\newcommand{\gtab}[1]{{\hypersetup{linkcolor=gesflanse}Table~\ref{#1}\hypersetup{linkcolor=blue}}}

\definecolor{Orange}{cmyk}{0,0.61,0.87,0}
\definecolor{JungleGreen}{cmyk}{0.99,0,0.52,0}
\definecolor{OliveGreen}{cmyk}{0.64,0,0.95,0.40}
\definecolor{Brown}{cmyk}{0,0.81,1,0.60}
\definecolor{RoyalBlue}{cmyk}{0.71,0.53,0,0.12}
\definecolor{Gray}{cmyk}{0,0,0,0.40}
\definecolor{LightPink}{cmyk}{0.0,0.25,0,0}
\definecolor{LLightPink}{cmyk}{0.0,0.10,0,0}
\definecolor{LightBlue}{cmyk}{0.25,0,0,0}
\definecolor{LightGray}{cmyk}{0,0,0,0.2}

\graphicspath{{figs/}{noteplot/}{plot/}}

\def\brc#1{\left(#1\right)}
\def\Re{\mathrm{Re}}
\def\GeV{\mathrm{GeV}}
\def\IncTheo[#1]#2{$\left({\color{Orange}#1},{\color{cyan}#2}\right)$}
\def\IncTheop[#1]#2{$\left({\color{Orange}#1},{\color{gray}#2}\right)$}
\def\IncTheopp[#1]#2{$\left(\,{\color{gray}#1},{\color{gray}#2}\right)$}

\begin{document}
\fontsize{12pt}{14pt}\selectfont

% \preprint{ACFI-T24-08}
% \nopagebreak

\title{%\begin{flushright}
       %\mbox{\normalsize IPMU18-xxxx}
       %\end{flushright}
			 %\vskip 20pt
    \vspace{-18mm}
      \hfill {\fontsize{12pt}{14pt}\selectfont ACFI-T24-08} \\[5mm]
       \textbf{\Large \titledef}} % Insert title here!!!

\author[1,2]{{\large Shao-Feng Ge} \footnote{\href{mailto:gesf@sjtu.edu.cn}{gesf@sjtu.edu.cn}}}
\author[3]{{\large Zhuoni Qian} \footnote{\href{mailto:zhuoniqian@hznu.edu.cn}{zhuoniqian@hznu.edu.cn}}}
\author[1,2,4]{{\large Michael J. Ramsey-Musolf} \footnote{\href{mailto:mjrm@sjtu.edu.cn}{mjrm@sjtu.edu.cn}}}
\author[4]{{\large Jia Zhou} \footnote{\href{mailto:jia@umass.edu}{jia@umass.edu}}}
\affil[1]{Tsung-Dao Lee Institute \& School of Physics and Astronomy, Shanghai Jiao Tong University, Shanghai 200240, China}
\affil[2]{Key Laboratory for Particle Astrophysics and Cosmology (MOE) \& Shanghai Key Laboratory for Particle Physics and Cosmology, Shanghai Jiao Tong University, Shanghai 200240, China}
\affil[3]{School of Physics, Hangzhou Normal University, Hangzhou, Zhejiang 311121, China}
\affil[4]{Amherst Center for Fundamental Interactions, Department of Physics, University of Massachusetts, Amherst, MA 01003, USA}

\date{\today}

\maketitle

\begin{abstract}
\fontsize{12pt}{14pt}\selectfont
We explore the prospects for probing new physics (NP) beyond the Standard Model (SM) at future lepton colliders through precision measurements of $e^+e^-\to f{\bar f}$ observables off the $Z$ resonance. We consider interference between SM contributions and those arising from dimension-6, four-fermion effective operators that encode the effects of NP, yielding a linear dependence on the latter. This linear
dependence in general increases with magnitude of the collision energy offset from the $Z$ pole. We consider a variety of asymmetries in order to enhance the NP-sensitivity while reducing experimental systematic and theoretical, SM uncertainties: an inclusive above and below $Z$-resonance total cross section asymmetry ($A_\sigma$) as well as the conventional forward-backward ($A_{\rm FB}$)  and polarization ($A_{\rm pol}$)  asymmetries. Based on projected statistical uncertainties at the Circular Electron-Positron Collider (CEPC), we find that t measurement of $A_\sigma$ could extend the sensitivity to the NP mass scale by as much as a factor of $\sim 7$ compared to the present reach obtained with the CERN Large Electron Positron Collider. Inclusion of projected systematic theoretical SM uncertainties substantially reduce this sensitivity gain. For $A_{\rm FB}$, inclusion of experimental systematic uncertainties has a marginal impact on the gain in NP reach, whereas SM theoretical uncertainties remain a significant barrier to realizing the full NP sensitivity. Analogous conclusions apply to the CERN Future Circular Collider (FCC-ee) and International Linear Collider (ILC). 

\end{abstract}

\section{Introduction}

Precision measurements at lepton colliders have played an essential 
role in probing the Standard Model (SM) of particle physics
\cite{ParticleDataGroup:2022pth}.
In particular, its renormalizability and associated
radiative corrections have been tested by various $Z$-pole precision
measurements at SLC and LEP \cite{ALEPH:2005ab}. 
In addition, precision measurements can also probe the
possible new physics (NP) beyond the SM. If the NP lies
sufficiently above the weak scale, one may characterize its
effects at low energy using effective operators.
For dimension-six (dim-6) operators, the LEP and SLC programs
yielded lower bounds on the NP scale of
$\mathcal{O}\brc{20}$\,TeV \cite{ALEPH:2006bhb, ALEPH:2010aa}. 

Future lepton colliders such as the
Circular Electron Positron Collider (CEPC) \cite{CEPCStudyGroup:2018ghi},
the Future Circular Collider with $e^+e^-$ (FCC-ee) \cite{FCC:2018evy},
and the International Linear Collider (ILC)
\cite{LCCPhysicsWorkingGroup:2019fvj} have much higher luminosities around
the $Z$ pole. To make it concrete, the ILC will produce at least
500 times greater luminosity in its GigaZ operation than the total integrated
luminosity at LEP and SLC while the number can even reach $(5\sim 8) \times 10^5$ times
at CEPC and FCC-ee \cite{Belloni:2022due}. Thus, the statistical uncertainty can be reduced
by at least two orders of magnitude at CEPC or FCC-ee to allow high
precision measurement of the $Z$ boson properties. 
For example, the uncertainties of $M_Z$ and $\Gamma_Z$ from the
$Z$ lineshape scan can be reduced to roughly 0.1\,MeV
\cite{CEPCPhysicsStudyGroup:2022uwl}.
In addition, measurements of the forward-backward and polarization asymmetries can
also yield a determination of the weak mixing angle $\sin^2\theta_W^\text{eff}$ with
unprecedented uncertainty of $10^{-5}$
at CEPC \cite{Zhao:2022lyl} and $10^{-6}$ at FCC-ee \cite{Bernardi:2022hny}.

In the context of dimension-6 effective operators
\cite{Buchmuller:1985jz,Grzadkowski:2010es},
the fit to the LEP-I and LEP-II data
\cite{Barbieri:1999tm,Han:2004az}
can probe the NP mass scale up to $\mathcal{O}\brc{10}$\,TeV.
With significantly improved electroweak precision measurements
at the future lepton colliders, one would expect enhanced NP sensitivity. For example, studies
considering probes of dimension-6 operators at CEPC, FCC-ee,
and ILC can be found in Refs.\,\cite{Ellis:2015sca,Ge:2016tmm,Durieux:2017rsg,Chiu:2017yrx,deBlas:2016ojx,deBlas:2016nqo}. 
Using {\tt HEPfit} \cite{DeBlas:2019ehy} for a
global fit to  SM electroweak precision observables
and taking into account present bounds on the NP corrections
\cite{deBlas:2016ojx,deBlas:2016nqo},
the sensitivity to the BSM physics scales improves by
one order of magnitude
at FCC-ee \cite{FCC:2018evy}. 
%In particular, the sensitivity
%of the dimension-6 four-fermion operator 
%$O_{\ell \ell} \equiv (\bar \ell \gamma_\mu \ell) (\bar \ell \gamma^\mu \ell)$
%improves from $\mathcal{O}\brc{7}$\,TeV to
%$\mathcal{O}\brc{46}$\,TeV \cite{deBlas:2016nqo} \mrmC{this is not what I see in their paper, and I don't understand their methodology}.

In this work, we analyze the sensitivity of future lepton colliders
to the dimension-6 four-fermion operators that contribute to the fermion pair
production $e^+ e^- \rightarrow f \bar f$. In particular,
we focus on the potential measurements off the $Z$ pole,
which admit sensitivity to interference between dimension-6 operator
and SM amplitudes, thereby yielding a linear dependence on the NP contributions.
We consider several types of asymmetries: 
the inclusive above- and below- $Z$-pole total cross section asymmetry, $A_\sigma$; the conventional forward-backward asymmetry $A_{\rm FB}$ at a single beam energy; and the beam polarization asymmetry, $A_\sigma$, also at a single beam energy.  In principle, these asymmetry measurement can yield enhanced NP sensitivity while reducing the impact on various experimental, systematic effects. 

Based on projected statistical uncertainties at the CEPC, we find that the a measurement of $A_\sigma$ could extend the sensitivity to the NP mass scale by as much as a factor of $\sim 7$ compared to the present reach obtained with the LEP. Inclusion of projected systematic  theoretical SM uncertainties substantially reduce this sensitivity gain. For $A_{\rm FB}$, inclusion of experimental systematic uncertainties has a marginal impact on the gain in NP reach, whereas SM theoretical uncertainties remain a significant barrier to realizing the full NP sensitivity. Analogous conclusions apply to the FCC-ee and ILC. 
Our results imply that realizing the full potential for NP sensitivity with electroweak precision measurements at future lepton colliders requires both further reductions in projected experimental systematic uncertainties and calculation of SM higher-order electroweak radiative corrections. 

Our paper is organized as follows. We first define in
\gsec{sec:observable} those
dimension-6 four-fermion operators that can contribute to the
$e^+ e^- \rightarrow f \bar f$ scattering process via
SM-NP interference off the $Z$ pole. The same section
also elaborates the asymmetry observables, the next leading
order (NLO) radiative corrections to the SM contributions,
the symmetrization and anti-symmetrization decomposition
of off-$Z$-pole observables. The uncertainties (including
statistical, theoretical, and experimental ones) and projected
sensitivities at future lepton colliders with CEPC as
illustration can be found in \gsec{sec:uncertainty}.
We further explore the possible improvement with beam
polarization and polarized asymmetry in \gsec{sec:polarization}.
The final \gsec{sec:conclusion} shows the combined sensitivities
and our conclusions. %Appendix \ref{app:eemumu} contains a compendium of detailed results for the $e^+e^-\to\mu^+\mu^-$ channel.

\section{Observing New Physics with Off-$Z$ Interference}
\label{sec:observable}

With good agreement between SM and existing collider searches,
new physics should appear at much higher energy than the $Z$
pole. If there is any new physics, its effect around $Z$ pole
should manifest itself as effective operators. Especially,
the $Z$ pole searches at future lepton colliders involves at
least two electron fields for the electron and positron
beams as initial states. In addition, the final state typically
involves two fermions, either two charged leptons or two
quarks. The lowest order of effective operators that can
accommodate these four fermion states is dimension six,
\begin{equation}
  \mathcal{L}_\text{eff}
=
  \mathcal{L}_{\rm SM}
+ \sum_i c_i \mathcal{O}_i, 
\end{equation}
where $\mathcal O_i$ is a set of dim-6
operators and $c_i$ is the corresponding coefficients
with dimension of inverse mass squared. Additional dimension-6
operators that can modify the $Zf\bar f$ couplings such as
$(\overline{Q}_i \gamma^{\mu} Q_i)(i H^{\dagger} \overset{\leftrightarrow}{D}_{\mu} H)$
can also contribute, but are constrained much better at the $Z$ pole.
The concrete four-fermion operators have been summarized in
\gtab{tbl:11opnew}.

\begin{table}[htb!]
\centering
\begin{tabular}{l|l}
  $\mu^+\mu^-$   & $q\bar{q}$ \\
\hline
  $\mathcal O^s_{LL}
\equiv
  \frac 1 2
  (\bar L \gamma^\mu L)
  (\bar L \gamma_\mu L)$
& $\mathcal O_{L Q}^s
\equiv
  (\bar{L}\gamma^\mu L) (\bar{Q}\gamma_\mu Q)$
\\
  $\mathcal O_{LL}^t
  = \frac 1 2 \brc{\bar{L}\gamma^\mu\sigma^a L} \brc{\bar{L}\gamma_\mu\sigma^a L}$
& $\mathcal O_{L Q}^t=\brc{\bar{L}\gamma^\mu\sigma^a L}\brc{\bar{Q}\gamma_\mu\sigma^a Q}$
\\
  $\mathcal O_{L \ell}=\brc{\bar{L}\gamma^\mu L}\brc{\bar{\ell}\gamma_\mu \ell}$
& $\mathcal O_{Q\ell} =\brc{\bar{Q}\gamma^\mu Q}\brc{\bar{\ell}\gamma_\mu \ell}$
\\
  $\mathcal O_{\ell\ell}=\frac{1}{2}\brc{\bar{\ell}\gamma^\mu \ell}\brc{\bar{\ell}\gamma_\mu \ell}$
& $\mathcal O_{L u} =\brc{\bar{L}\gamma^\mu L}\brc{\bar{q}_u\gamma_\mu q_u}$
\\
& $\mathcal O_{L d} =\brc{\bar{L}\gamma^\mu L}\brc{\bar{q}_d\gamma_\mu q_d}$
\\
& $\mathcal O_{\ell u} =\brc{\bar{\ell}\gamma^\mu \ell}\brc{\bar{q}_u\gamma_\mu q_u}$
\\
& $\mathcal O_{\ell d} =\brc{\bar{\ell}\gamma^\mu \ell}\brc{\bar{q}_d\gamma_\mu q_d}$
\end{tabular}
\caption{The dimension-6 four-fermion operators that can interfere with
the SM contributions to the $e^+ e^- \rightarrow f \bar f$ process \cite{Han:2004az}.}
\label{tbl:11opnew}
\end{table}

These dimension-6 four-fermi
operators are divided into two groups according
to the final state of the $e^+ e^-$ collision.
For charged leptons, the $e^+ e^- \rightarrow \mu^+ \mu^-$
leaves the most clear signal in the detector.
Requiring the SM $SU(2)_L \times U(1)_Y$ symmetries,
the lepton bilinear can be constructed in terms of
either left-handed doublets ($L$) or right-handed
singlets ($\ell$). Being doublet, the left-handed
leptons can form either $SU(2)_L$ singlet or triplet
where the latter has a $\sigma^a$ matrix.
So the purely left-handed operators have singlet
($\mathcal O^s_{LL}$) and triplet ($\mathcal O^t_{LL}$)
forms while those involving right-handed leptons
have only singlet forms ($\mathcal O_{L \ell}$)
and ($\mathcal O_{\ell \ell}$). We have omitted
the $s$ subscript for simplicity. For hadronic
final states, $e^+ e^- \rightarrow j j$ with
$j \equiv udscb$, the operators contain one lepton
bilinear and a quark one. While $L$ ($Q$) denotes
the left-handed lepton (quark) field, $\ell$ ($q_{u, d}$
for up and down type quarks)
denotes the right-handed components.

As elaborated in
\gsec{eq:interference}, we consider only
vector currents such that the dimension-6 operator
can interfere with the SM contributions to
make its effect already appear at order
$\mathcal{O}(1/\Lambda^2)$. For comparison,
the scalar and tensor bilinears involve both left-
and right-handed fermions. It is possible to
also construct scalar operators like
$(\bar L \ell) (\bar \ell L)$ where the two
scalar bilinears $\bar L \ell$ and $\bar \ell L$
are $SU(2)_L$ doublets. Note that the initial
$e^+ e^-$ can be contributed by $\bar L$ and
$L$ ($\bar \ell$ and $\ell$) while the final
state by $\bar \ell$ and $\ell$ ($\bar L$ and $L$)
which can still have interference with the
SM contributions. Nevertheless, the $\bar L \ell$
and $\bar \ell L$ bilinears indicate that the
scalar mediator has flavor changing Yukawa couplings.
For simplicity, we also impose flavor conservation
to forbid such scalar and tensor operators.
The combination of interference and flavor
conservation significantly reduces the number
of four-fermion operators. In addition, flavor
universality and lepton number conservation are
also assumed.

\begin{table}[htbp]
\centering
\begin{tabular}{c|c|*{11}{c}}
\multicolumn{2}{c|}{} & $\mathcal O_{LL}^s$ & $\mathcal O_{LL}^t$ & $\mathcal O_{L\ell}$ & $\mathcal O_{\ell\ell}$ & $\mathcal O_{LQ}^s$ & $\mathcal O_{LQ}^t$  & $\mathcal O_{Q\ell}$ & $\mathcal O_{L u}$ & $\mathcal O_{L d}$ & $\mathcal O_{\ell u}$ & $\mathcal O_{\ell d}$ \\
%\hline
%\multirow{2}{*}{best-fit} &$c_i$ &$-5.0$ &$-5.8$ &$-60.$ &$-6.9$ &$-0.3$ &$-23.$ &$\gred{-4.1~10^2}$ &$\gred{-78.}$ &$7.5$ &$\gred{-5.9~10^2}$ &\gred{$-6.4~10^2}$ \\
%&$\Lambda_i$/TeV &15.6 &14.7 &4.58 &13.5 &64.7 &7.4 &1.8 &4.0 &12.9 & 1.5 & 1.4 \\
\hline
\multirow{2}{*}{95\% CL} &$|c_i|^{\rm max}$ &$9.2$ &$1.3$ &$5.5$ &$10.1$ &$4.3$ &$5.5$ &$4.5$ &$7.8$ &$8.3$ &$9.0$ &$6.7$ \\
&$\Lambda_i^{\rm min}$/TeV &11.7 &30.7 &15.0 &11.2 &17.1 &15.1 &16.8 &12.7 &12.3 &11.8 & 13.7
\end{tabular}
\caption{%The best-fit values for the Wilson coefficients $c_i$ of the 11 dimension-6 four-fermion operators in the unit of $10^{-8}\GeV^{-2}$.
The allowed maximum of Wilson coefficients $|c_i|^{\rm max}$ and corresponding NP scales $\Lambda_i^{\rm min} \equiv \sqrt{4\pi/\left|c_i\right|}$ in the unit of $10^{-8}\GeV^{-2}$.}
\label{tbl:coe}
\end{table}
The operator coefficient $c_i$ can also be parametrized in
terms of the corresponding NP scale
$\Lambda_i$, with $\left|c_i\right| \equiv 4 \pi/\Lambda_i^2$.
Current bounds on the set of dim-6 operators are obtained from a $\chi^2$ fit with uncertainties and best fit values calculated from
Eq.(A1) and coefficients in Tables III \& IV of 
\cite{Han:2004az}. By implementing one operator at a time, the allowed largest value of the Wilson coefficients $|c_i|^{\rm max}$ and the corresponding NP scale $\Lambda_i^{\rm min}$ are calculated and listed in \gtab{tbl:coe}.
The bounds include data from the LEP-I $Z$-pole scan,
the LEP-II run with energy range up to 200 GeV,
the $Z$-pole scan at SLD with polarized beams, and the
neutrino-nucleus scattering measurements.

\subsection{Manifesting New Physics with Interfence around the $Z$ Pole}
\label{eq:interference}

We study the processes $e^-e^+\to \mu^-\mu^+$ and $e^-e^+\to q\bar q$
($q=u,d,c,s,b$) near the $Z$-pole region. Both SM with an $s$-channel
$Z/\gamma$ mediator and new physics via the four-fermion opearators
in \gtab{tbl:11opnew} can contribute,
\begin{subequations}
\begin{align}
\hspace{-2mm}
  \mathcal M_{\rm SM}
& =
  \frac {\bar u_f \gamma_\mu (g_{fL} P_L + g_{fR} P_R) v_f
        i
        \bar v_e \gamma^\mu (g_{eL} P_L + g_{eR} P_R) u_e}
        {s - M^2_Z + i M_Z \Gamma_Z}
+
  i \frac {Q_f Q_e e^2} s
  (\bar u_f \gamma_\mu v_f)
  (\bar v_e \gamma^\mu u_e),
\label{eq:Msm}
\\
\hspace{-2mm}
  \mathcal M_{\mathcal O_i}
& =
  F \times i c_i
  (\bar u^i_f \gamma_\mu v^i_f)
  (\bar v^i_e \gamma^\mu u^i_e).
  \label{eq:amp:new}
\end{align}
\end{subequations}
Of the SM amplitude $\mathcal M_{\rm SM}
\equiv \mathcal{M}_{\rm SM}^Z + \mathcal{M}_{\rm SM}^\gamma$ in
\geqn{eq:Msm}, the first term is the $Z$ ($\mathcal{M}_{\rm SM}^Z$)
and the second the photon ($\mathcal{M}_{\rm SM}^\gamma$)
mediated contributions. The $Z$ pole
line shape is captured by the $Z$ mass $M_Z$ and its decay
with $\Gamma_Z$.
The spinor $u^i_{f,e}$ ($v^i_{f,e}$)
denotes the
left- and right-handed final-state fermion $f \equiv \mu, q$
(anti-fermion)
and initial-state electron (positron).
For generality, we use $g_{fL,R}$ and
$g_{eL,R}$ to denote the coupling of left- and right-handed
fermion and electron with the $Z$ gauge boson.
In SM, these couplings are, 
\begin{equation}
  g_{fL}
\equiv
  \frac g {c_W} (T_{W,f}^3-Q_fs_W^2),
\quad \mbox{and} \quad
  g_{fR}
\equiv
- \frac g {c_W} Q_f s^2_W,
\label{eq:chiral:cpl}
\end{equation}
where $Q_{e(f)}$ is the electric charge of the
initial/final-state fermions. Some new physics may
enter as correction to the $Z$ couplings, such as
oblique corrections, to modify $g_{fL}$ and $g_{fR}$.
For simplicity, we omit such corrections and
focus on the four-fermion operators.

For the new physics contribution
$\mathcal M_{\mathcal O_i}$,
the prefactor $F$ is 1 except for the following cases,
$F=1/2$ for $\mathcal{O}_{LL}^{s,t}$
and $\mathcal O_{\ell\ell}$ operators while
$F=-1$ for the $\mathcal{O}_{L Q}^t$ operator when the
final states are up-type quarks. Note that two kinds of
$\mathcal{O}_{L\ell}$ operators contribute to the
$e^-e^+\to \mu^-\mu^+$ process with $L=\mu$ and $\ell=e$
or $L=e$ and $\ell=\mu$. 

The differential cross section for the fermion pair production
$e^+e^-\rightarrow f\bar{f}$ contains the purely SM contributions
$\propto|\mathcal M^Z_{\rm SM} + \mathcal M^\gamma_{\rm SM}|^2$,
the dimension-6 operator contribution $\propto |\mathcal M_{\mathcal O_i}|^2$,
as well as the interference between them. While the purely
new physics term $|\mathcal M_{\mathcal O_i}|^2$ is order
$1/\Lambda^4$, the interference term
$\mathcal M_{\mathcal O_i} \sim 1/\Lambda^2$ is less suppressed.
It is reasonable to expand as power series of $1/\Lambda^2$.
Up to the linear term of $1/\Lambda^2$, the four-fermion operator
contribution is,
\begin{equation}
  d\sigma_{\mathcal O_i}
\approx
  \frac 1 {2 s}
  \sum_f \frac {N_f} 4
    2\Re\left[\mathcal{M}_{\mathcal O_i}\cdot
    \brc{\mathcal{M}_\mathrm{SM}^Z 
   +
   \mathcal{M}_\mathrm{SM}^\gamma}^*
   \right]
   d\Phi_2,
%\end{aligned}
\label{eq:dsec}
\end{equation}
where $1/4$ is from averaging over
the initial-state spins and $N_f$
the color degree of freedom
of the final states with $N_f = 1$ for leptons and
$N_f = 3$ for quarks, respectively. 
The SM and interference contributions can be analytically obtained
using various tools such as {\tt FORM}~\cite{Vermaseren:2000nd,Kuipers:2012rf}, {\tt Package-X}~\cite{Patel:2015tea,Patel:2016fam}, and {\tt FeynCalc}~\cite{Mertig:1990an,Shtabovenko:2016sxi,Shtabovenko:2020gxv}.

Integrating (\ref{eq:dsec}) over the two body phase space $d \Phi_2$, one obtains the analytical expression for the leading order total cross sections 
\cite{Altarelli:1989hv},
%\jzR{add reference here; the expressions could be compared with that in P.~93 (2.17)},
%\gesfC{Ref here for the SM cross sections to make consistency check?}
\begin{subequations}
\begin{eqnarray}
  \sigma^Z_{\rm SM}
& = &
  \sum_f \frac {N_f}{48 \pi}
  \frac{\brc{g_{eL}^2 + g_{eR}^2} 
         \brc{g_{fL}^2 + g_{fR}^2}s}
       {\brc{s-M_Z^2}^2 + \Gamma_Z^2 M_Z^2}, 
\label{eq:xsec:zz}
\\
  \sigma^{Z \gamma}_{\rm SM}
& = &
  \sum_f \frac {N_f}{48 \pi}
  \frac{2 (4 \pi \alpha) Q_e Q_f 
        \brc{g_{eL} + g_{fR}}
        \brc{g_{fL} + g_{fR}}
        \brc{s-M_Z^2}}
       {\brc{s-M_Z^2}^2+\Gamma_Z^2M_Z^2}, 
\label{eq:xsec:zg}
\\
  \sigma^\gamma_{\rm SM}
& = &
  \sum_f \frac {N_f}{48 \pi}
  \frac{4 (4 \pi \alpha)^2 Q_e^2 Q_f^2} s,
\label{eq:xsec:gg}
\\
  \sigma_{\mathcal{O}_i}^{\rho\lambda}
& = &
  \sum_f F~\frac {N_f}{4 8 \pi}
  2 c_i 
\left[
  \frac{g_{e \rho}g_{f \lambda} s \brc{s-M_Z^2}}{\brc{s-M_Z^2}^2+\Gamma_Z^2M_Z^2}
+ 4 \pi \alpha Q_e Q_f
\right].
\label{eq:xsec:d6}
\end{eqnarray}
\label{eq:xsecs}
\end{subequations}
The inclusive measurement sum over all quark flavors
with $\sum_f$ except for the top. Of the two terms in
\geqn{eq:xsec:d6}, the first comes from the interference
of $\mathcal M_{\mathcal O_i}$ with the $Z$ mediated
contribution $\mathcal M^Z_{\rm SM}$ and the second
with the photon one $\mathcal M^\gamma_{\rm SM}$.
The superscripts $\rho$ and $\lambda$ in
$\sigma_{\mathcal{O}_i}^{\rho\lambda}$ are the chiralities
of the initial- and final-state particles, i.e., 
$\rho,~\lambda=L,R$.
We write the cross sections with symbolic chiral couplings
so that each contribution with particular initial- and
final-state chiralities can be easily extracted.
From these, one may readily obtain the initial-state
polarized cross sections whose consequences would
be elaborated in \gsec{sec:polarization}. Before that
we mainly consider unpolarized beams.
One may also refer to the general
description of fermion pair production above the $Z$
pole in the Sec.\,2.6.1 of \cite{Hollik:2000ap} where
the analytical formula for the total and FB
cross section is in the Eqs.(1.137) and (1.138).

In order to evaluate the cross sections and event rates,
it is necessary to input the model parameters in the
electroweak (EW) sector, including the two gauge couplings
$g_1$ and $g_2$, the weak scale or equivalently the
Higgs vacuum expectation value $v$, and the $Z$ line
shape parameter or equivalent the $Z$ decay width $\Gamma_Z$,
\begin{subequations}
\begin{align}
  G_F
& =1.16637\times10^{-5}\,\GeV^{-2},
&
  \alpha^{-1} \brc{M_Z} = 128.946, 
\\
  M_Z
& = 91.1876\,\GeV, 
&
  \Gamma_Z = 2.4952\,\GeV, 
\end{align}
\end{subequations}
Besides $\Gamma_Z$, the other three can be inferred
from the fine-structure constant $\alpha$, the Fermi
constant $G_F$, and the $Z$ mass $M_Z$. This set
$(\alpha, G_F, M_Z)$ of parameters are the most
precisely measured EW parameters. Different sets
of EW input parameters for SM Effective Field Theory
(SMEFT) predictions at the LHC are recently discussed
\cite{Brivio:2021yjb}.

\subsection{NLO Radiative Corrections to the SM Contributions}

Higher-order EW radiative corrections to the fermion pair
production amplitude can significantly modify the predicted
$Z$ lineshape. In particular, the QED photonic corrections that could
amount to about 30\% near the $Z$ pole \cite{ALEPH:2005ab}.
It involves the photon
initial-state radiation (ISR), final-state radiation (FSR), and
their interference (IFI). Of them, the ISR contribution
has the largest influence. 
While the photon ISR is highly suppressed right at the $Z$ pole, its relative
contribution increases with the distance away from the $Z$ pole.
Therefore, one has to include these photonic effects. 
An analytic calculation to the complete set of the QED corrections
including soft photon exponentiation to order $\mathcal{O}\brc{\alpha}$
for $e^+e^-\to f\bar{f}$ was studied in Ref.\,\cite{Bardin:1990fu}.

A variety of codes are available for the full EW radiative corrections
to the $Z$ production at $e^+e^-$ colliders. For example,
{\tt TOPAZ0} \cite{Montagna:1993py,Montagna:1993ai,Montagna:1998kp}
and {\tt ZFITTER} \cite{Bardin:1999yd,Arbuzov:2005ma} are semi-analytical
{\tt Fortran} programs while {\tt $\mathcal{KK}$MC}
\cite{Jadach:1999vf,Jadach:2000ir} is a Monte Carlo event generator.
In what follows, we use {\tt ZFITTER} to implement the EW
radiative corrections, including one-loop corrections and
real photon radiations (ISR, FSR \& IFI), for the total cross sections
as well as the polarized and forward-backward (FB) asymmetries.

\subsection{Symmetrization and Anti-Symmetrization}
\label{sec:symmetrization}

From the total cross section formulas in \geqn{eq:xsecs},
one may see that the new physics contribution
$\sigma_{\mathcal{O}_i}^{\rho\lambda}$ has apparent
asymmetric feature around the $Z$ pole. This is
in particularly true for the interference term with the
SM $Z$ contribution $\mathcal M^Z_{\rm SM}$, namely,
the first term in the bracket of \geqn{eq:xsec:d6}.
The new physics contribution not just reverses sign
with the distance $s - M^2_Z$ from the $Z$ pole,
but also increases with $|s - M^2_Z|$. It is of
great advantages to use the off-$Z$-pole run of
future lepton colliders to search for new physics
that enters as four-fermion operators.

\begin{figure}[t]
\centering
\includegraphics[width=0.45\textwidth]{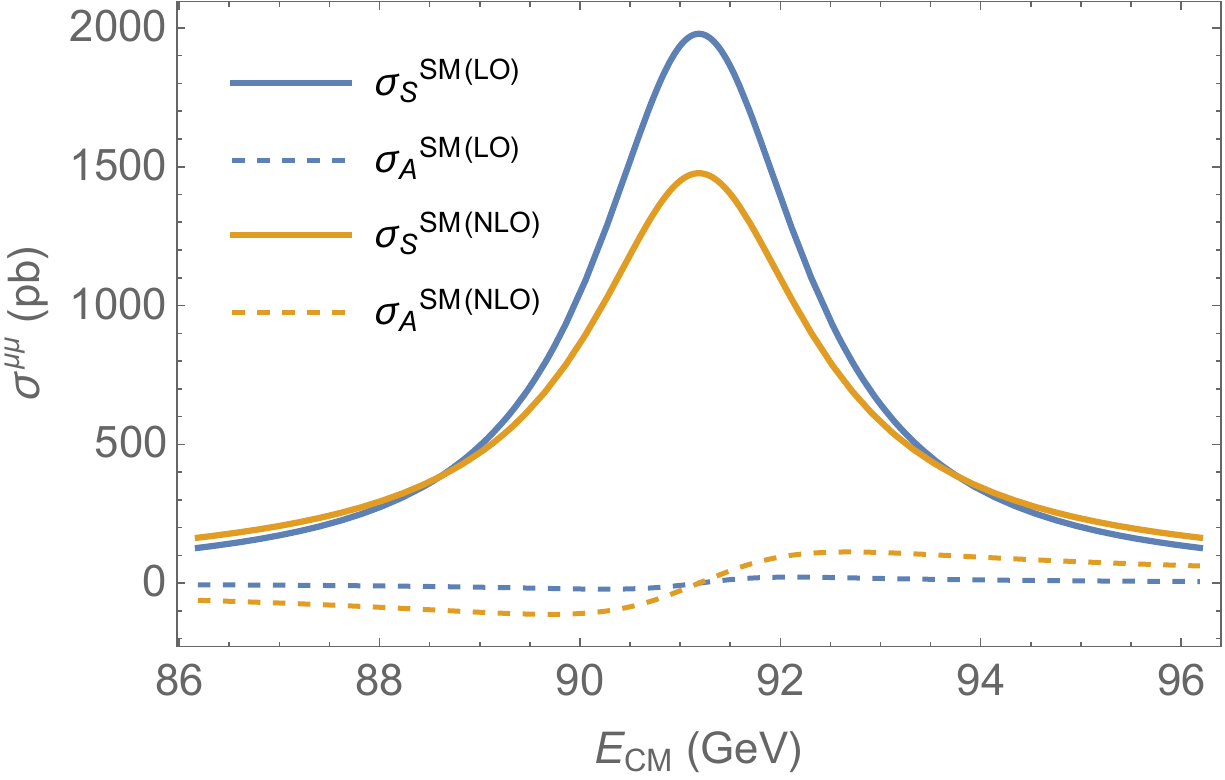} \hspace{0.4cm}
\includegraphics[width=0.46\textwidth]{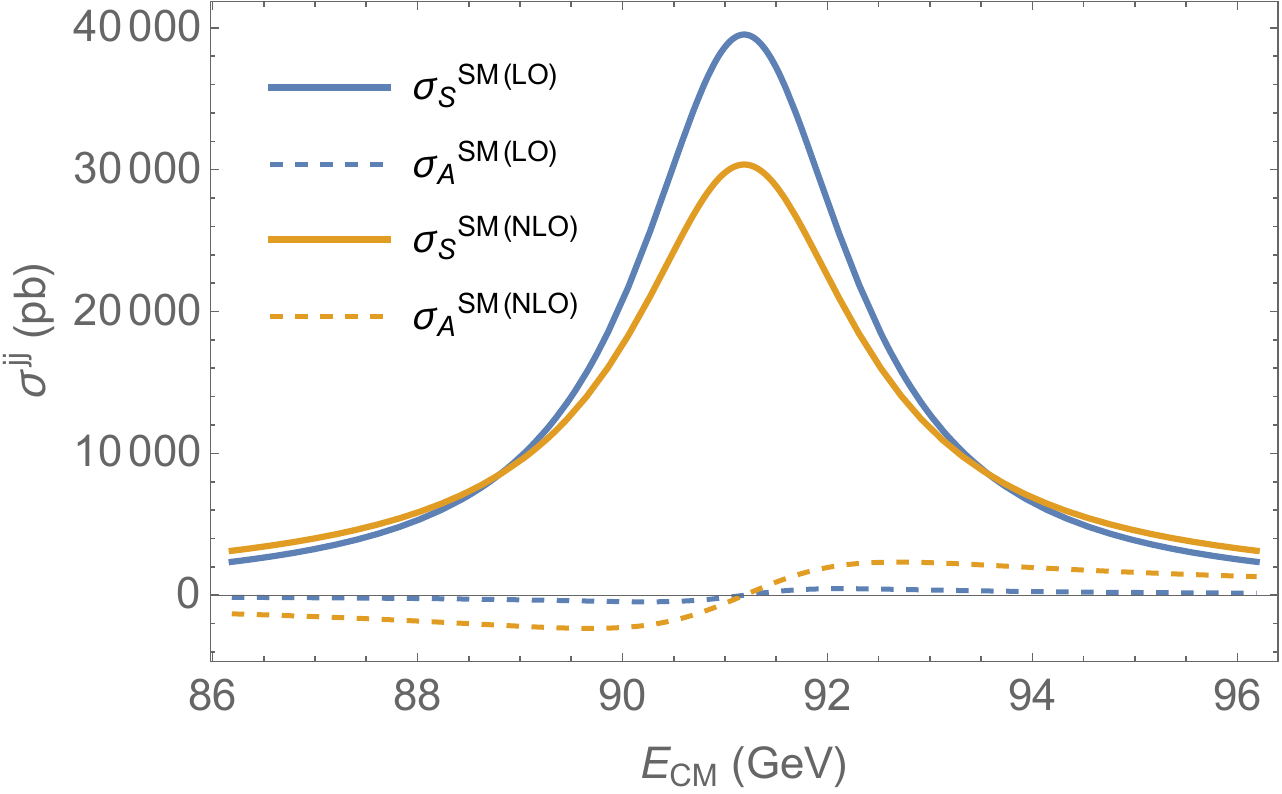} 
\\[4mm]
\includegraphics[width=0.48\textwidth]{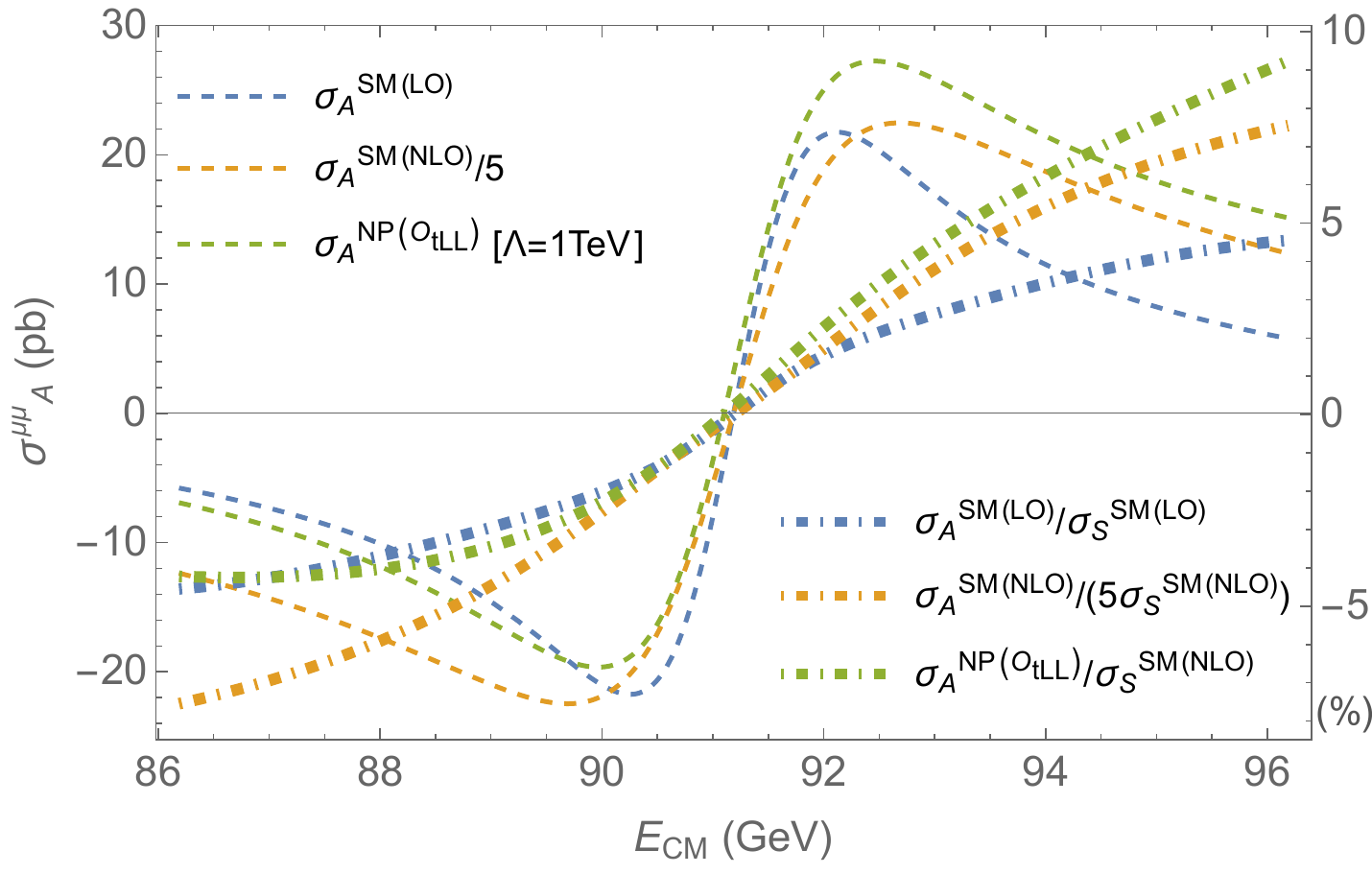}
\includegraphics[width=0.49\textwidth]{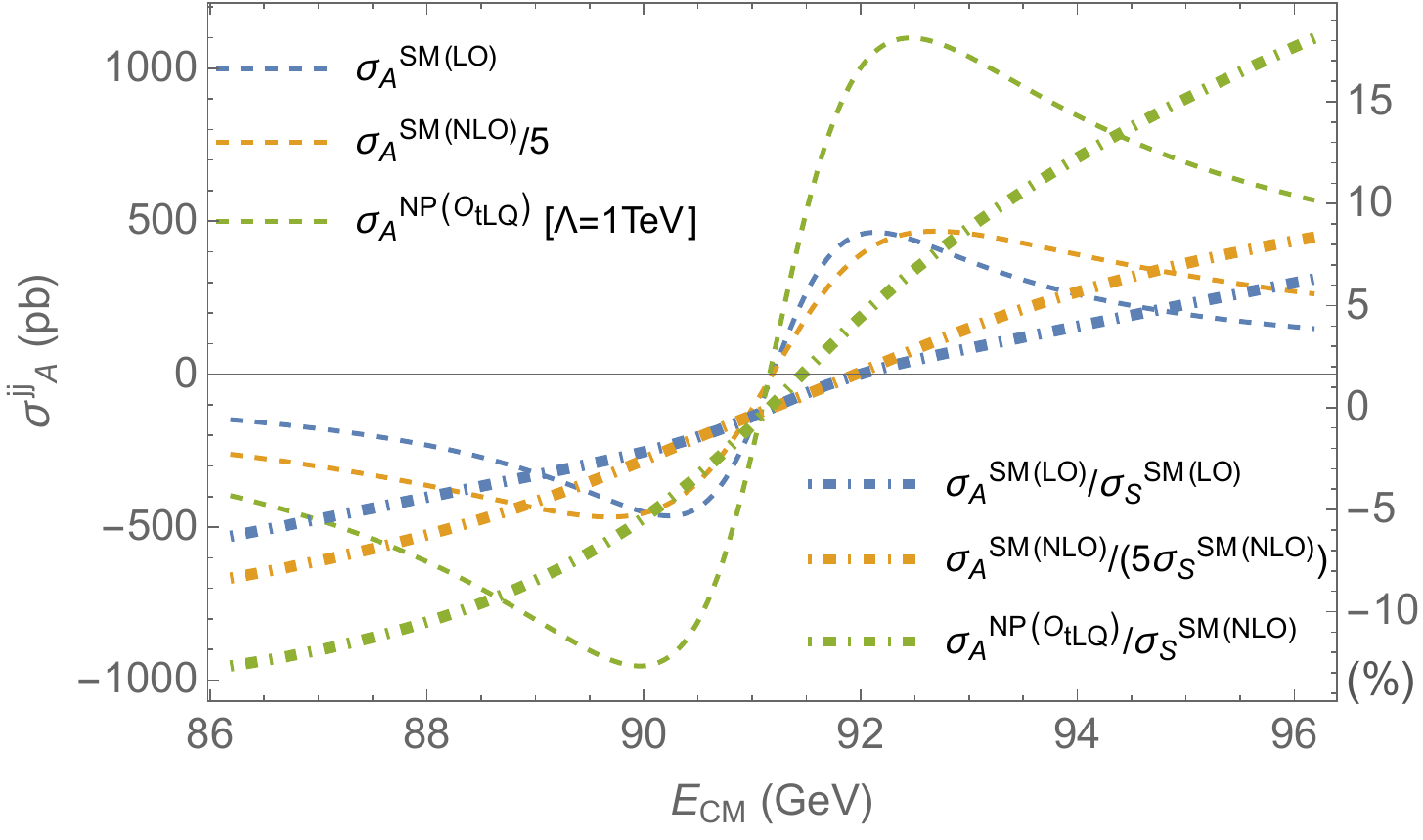}
\caption{The symmetric ($\sigma_S$) and anti-symmetric ($\sigma_A$)
parts of the total cross sections for the $e^+ e^- \rightarrow \mu^+ \mu^-$
(Left) and $e^-e^+\to q\bar q$ (Right) processes. While the upper
panels show the comparison with the SM contributions, the
lower panels show the absolute (dashed) and relative
(thick dot-dashed) size of the anti-symmetric
contributions from the SM and new physics. Due to its much larger size,  $\sigma_A^{\rm SM(NLO)}$ are scaled down by five times. 
}
\label{fig:symmetrization}
\end{figure}

It is then necessary to invent a way to see the
symmetric and anti-symmetric part of the total cross
sections around the $Z$ pole. Considering a region
of $\sqrt s = M_Z \pm \Delta$ with $\Delta \le 5$\,GeV),
the total cross section for $e^-e^+\to f \bar f$ can
be decomposed into symmetric and anti-symmetric parts,
\begin{equation}
  \sigma_{S,A}(\sqrt s)
\equiv
  \frac 1 2
\left[
  \sigma_{tot}(\sqrt s)
\pm
  \sigma_{tot}(2 M_Z - \sqrt s)
\right].
\end{equation}
The symmetric and anti-symmetric parts of the SM and new
physics contributions have been shown in \gfig{fig:symmetrization}
for comparison.
One may see that even the SM contributions have both
symmetric (solid) and anti-symmetric (dashed) parts as
shown in the upper panels of \gfig{fig:symmetrization}.
It is interesting to observe that the anti-symmetric
parts are always positive above the $Z$ pole ($\sqrt s > M_Z$)
with $s - M^2_Z$ dependence in the numerators of
\geqn{eq:xsec:zg} and \geqn{eq:xsec:d6}.
The major SM contribution from the $Z$ mediation is symmetric
while the anti-symmetric part at leading order (LO) comes
from the $Z$-$\gamma$ interference $\sigma^{Z \gamma}_{\rm SM}$
as summarized in \geqn{eq:xsec:zg}. However, the NLO
radiative correction contributes much larger anti-symmetric
components. As shown in the lower panels of
\gfig{fig:symmetrization}, the NLO anti-symmetric part
is several times of its SM LO counterpart and the new
physics one with $\Lambda = 1$\,TeV. It is necessary to
incorporate the SM NLO radiative corrections when studying
the four-fermion operators with off-$Z$-pole run.

The anti-symmetric parts in the lower panels of
\gfig{fig:symmetrization} vanishes at the $Z$ pole
with $s - M^2_Z = 0$. When moving away from the $Z$ pole,
their sizes first increase with the distance $|s - M^2_Z|$
and reaches the maximum values with an offset around
$|\sqrt s - M_Z| \approx 1$\,GeV
before starting to decrease. Around the deviation around
1\,GeV, both SM and new physics predicts the largest
asymmetries with a relative size reaching around 35\% and
7\% with $\Lambda_i = 1$\,TeV as benchmark, respectively.
So a reasonable off-$Z$-pole run scheme should have
$\mathcal O($GeV) offset. Not too far and not too close.

Currently, the CEPC already has a $Z$ lineshape scan scheme
with five energy points around the $Z$ pole. The typical
offsets are 1\,GeV and 3\,GeV, respectively, which fits
the off-$Z$-pole run for the four-fermion new physics
searches. 
\begin{table}[h]
\centering
\begin{tabular}{c|ccccc}
  $\sqrt{s}$ (GeV) & 87.9 & 90.2 & 91.2 & 92.2 & 94.3 \\
\hline
% Luminosity (ab$^{-1}$) & 0.25 & 0.25 & 7 & 0.25 & 0.25
  Luminosity (ab$^{-1}$) & 1 & 1 & 100 & 1 & 1
\end{tabular}
\caption{ The updated CEPC snowmass report in 2021 projects a total integrated luminosity of $100$ ab$^{-1}$ in the vicinity of the $Z$-pole, yielding about $4\times$ 10$^{12}$ $Z$ boson. The corresponding off-peak integrated luminosity projection is 1 ab$^{-1}$. \cite{CEPCPhysicsStudyGroup:2022uwl} 
}.
\label{tbl:cepc:zscan}
\end{table}

For quantitative evaluation of the asymmetry to extract
the new physics contribution, we define the off-$Z$-pole
cross section asymmetry $A^i_\sigma$ ($i=\{\mu,q\}$),
\begin{equation}
   A_\sigma (\Delta_\pm)
\equiv
  \frac {\sigma (M_Z + \Delta_+) - \sigma (M_Z - \Delta_-)}
        {\sigma (M_Z + \Delta_+) + \sigma (M_Z - \Delta_-)}, 
\label{eq:xsec:asym:gen}
\end{equation}
between two running points,
$\sqrt{s_\pm} \equiv M_Z + \Delta_\pm$, where $\Delta_\pm$
are the energy offsets above and below the $Z$ pole,
respectively. In principle, the two offsets $\Delta_\pm$
can be different. For simplicity, we take the same
offset, $\Delta = \Delta_\pm$.

\begin{figure}[t]
\centering
\includegraphics[width=0.48\textwidth]{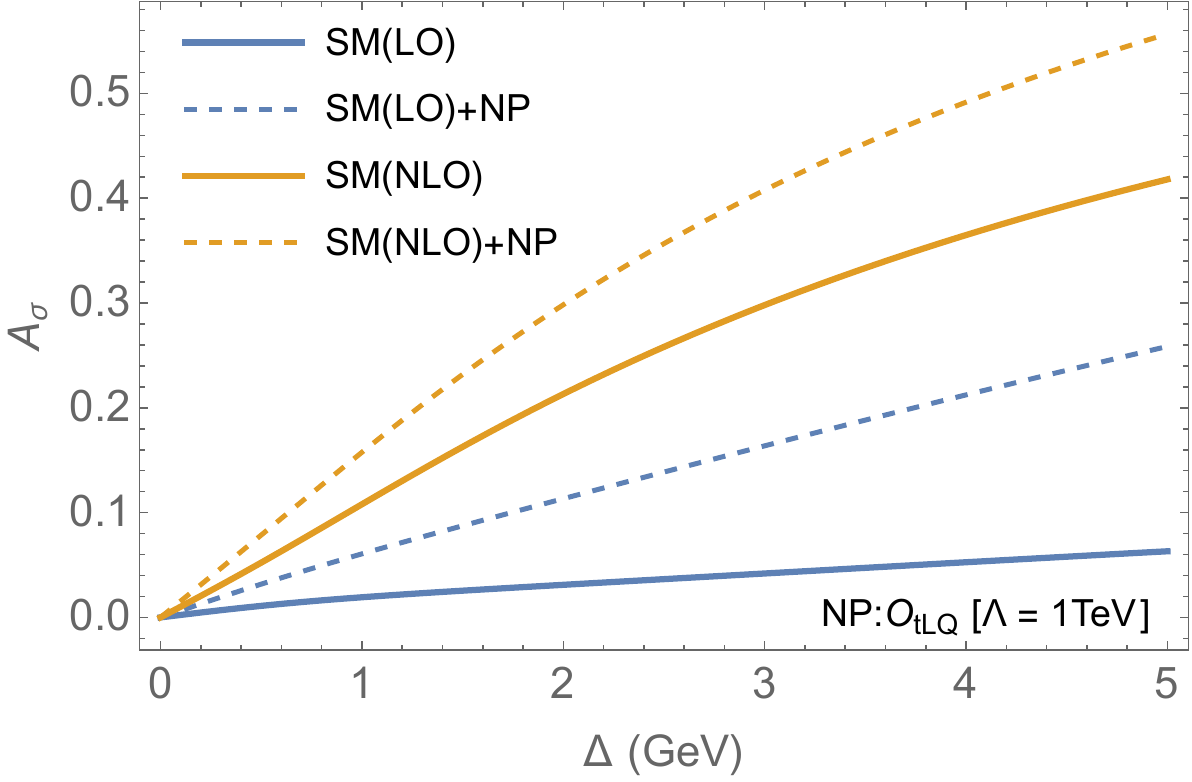}
\includegraphics[width=0.48\textwidth]{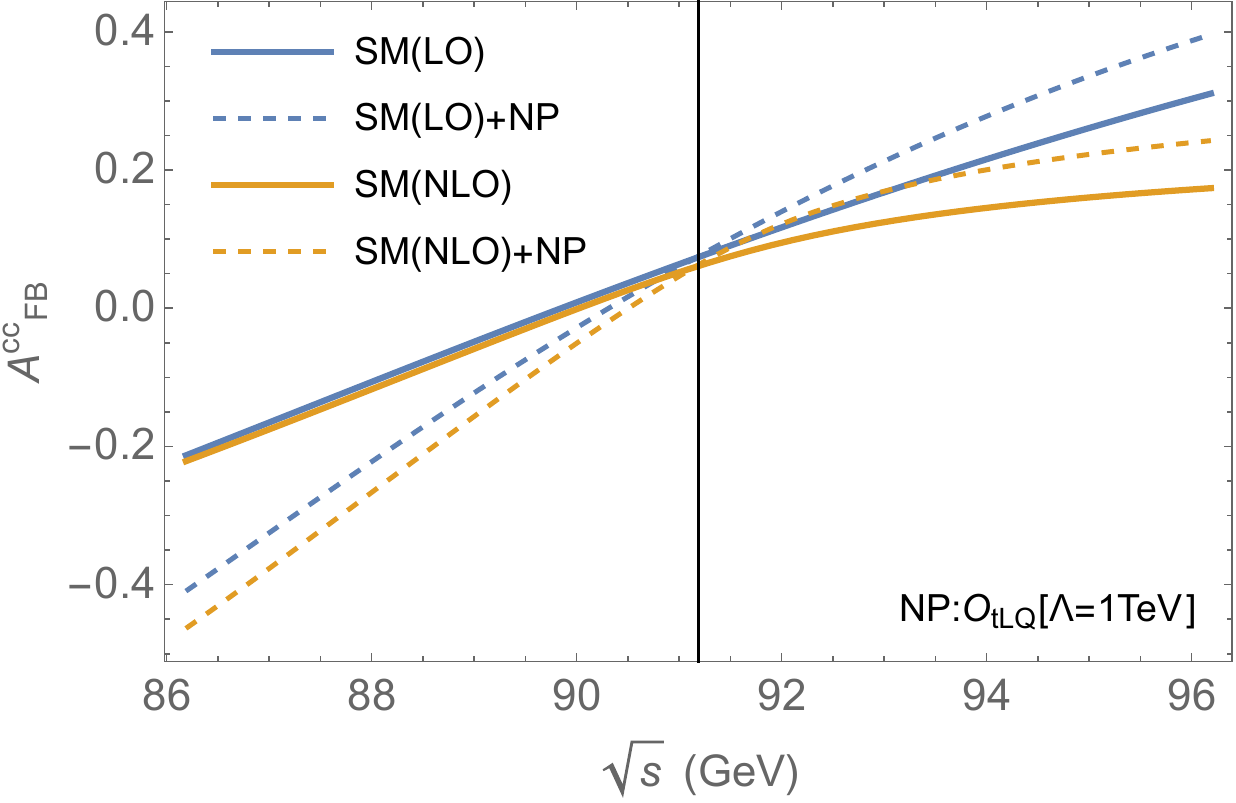}
\caption{The inclusive $A^q_\sigma$ (Left) and forward-backward
$A^c_{\rm FB}$ (Right) asymmetries from both SM (solid)
and the SM+NP four-fermion operator $\mathcal O^t_{LQ}$ (dashed)
with $\Lambda = 1$\,TeV. The effect of SM LO (blue)
and NLO (yellow) calculations are also shown for comparison.}
\label{fig:obs-lo-nlo}
\end{figure}

As pointed out above, both SM and four-fermion operators
can have anti-symmetric contributions.
\gfig{fig:obs-lo-nlo} shows the SM contribution $A^{\rm SM}_\sigma$
at both LO (blue) and NLO (yellow). The SM prediction increases
quite significantly after including the NLO radiative corrections.
For comparison, the new physics contribution via four-fermion
operators with $\Lambda = 1$\,TeV is relatively smaller but still
at the same order.
An indication of new physics occurs when
$A^{\rm NP}_\sigma$ exceeds the uncertainty in the
experimental measurement of $A_\sigma$. Note that the asymmetry increases with the offset $\Delta$ which is not necessarily the case for all operators and observable. 

\subsection{Forward-Backward Asymmetry}

As elaborated in \gsec{sec:uncertainty}, the precision
measurement around the $Z$ pole has large enough statistics
but suffers from experimental uncertanties, such as
the luminosity and energy uncertainties. Especially,
the cross section asymmetry measurement involves
two energy points that can have uncorrelated uncertainties
that cannot cancel out and will cause a huge problem.

The forward-backward asymmetry defined at a single
energy point has the benefit of allowing the luminosity
uncertainties to cancel. To be concrete, the forward-backward
asymmetry $A_{\rm FB}$,
\begin{equation}
    A_{\rm FB}(\sqrt s)
\equiv
  \frac {\sigma_{\rm F}(\sqrt s) - \sigma_{\rm B}(\sqrt s)}
        {\sigma_{\rm F}(\sqrt s) + \sigma_{\rm B}(\sqrt s)},
\label{eq:AFB}
\end{equation}
is defined as the asymmetry between the forward ($\sigma_{\rm F}$)
and backward ($\sigma_{\rm B}$) cross sections. The forward
(backward) events have negatively charged final-state fermion
moving in the same direction as the initial-state electron.
In other words, the scattering angle $\theta$ between the
negatively charged final-state fermion and the initial electron
should be smaller than $90^\circ$. Given the fact that flavor and charge identification  
for muon, bottom quark, and charm quark is welll studied at the lepton collider \footnote{Jet flavor and charge identification is ongoing and fastly improving study at the CEPC \cite{Liang:2023yyi}. The strange quark is foreseeable to contribute with slightly lower flavor tagging rate and reasonable charge determination}, we consider and combine forward-backward asymmetry measurement
from these three cases of final states in our analysis.

The FB cross sections can be obtained by integrating
over the scattering angle $\theta$ in two regions:
$(0,\pi/2)$ (forward) and $(\pi/2,\pi)$ (backward),
\begin{subequations}
\begin{align}
\hspace{-3mm}
  \sigma_{{\rm{SM}}, F}
& =
    \sum_f N_f
    \Bigg\{
    \frac{1}{384\pi}\frac{\left[g_{eL}^2\brc{7g_{fL}^2+g_{fR}^2}+g_{eR}^2\brc{g_{fL}^2+7g_{fR}^2}\right]s}{\brc{s-M_Z^2}^2+\Gamma_Z^2M_Z^2}
\nonumber
\\
& +
  \frac{\alpha}{48}\frac{Q_e Q_f \left[g_{eL}\brc{7 g_{fL}+g_{fR}}+g_{eR}\brc{g_{fL}+7g_{fR}}\right]\brc{s-M_Z^2}}{\brc{s-M_Z^2}^2+\Gamma_Z^2M_Z^2} 
    +\frac{2\pi\alpha^2}{3}\frac{Q_e^2Q_f^2}{s}
    \Bigg\},
\\
\hspace{-3mm}
    \sigma_{{\rm{SM}}, B}
& =
    \sum_f N_f
    \Bigg\{
    \frac{1}{384\pi}\frac{\left[g_{eL}^2 \brc{g_{fL}^2+7g_{fR}^2}+g_{eR}^2\brc{7g_{fL}^2+g_{fR}^2}\right]s}{\brc{s-M_Z^2}^2+\Gamma_Z^2M_Z^2}
\nonumber
\\
& +
  \frac{\alpha}{48}\frac{Q_e Q_f \left[g_{eL} \brc{g_{fL}+7g_{fR}}+g_{eR}\brc{7g_{fL}+g_{fR}}\right]\brc{s-M_Z^2}}{\brc{s-M_Z^2}^2+\Gamma_Z^2M_Z^2} 
    +\frac{2\pi\alpha^2}{3}\frac{Q_e^2Q_f^2}{s}
    \Bigg\},
\\
\hspace{-3mm}
  \sigma_{\mathcal{O}_i,F(B)}^{\rho\lambda}
& =
  \sum_f F \times c_i \frac{N_f\alpha}{48} s
\left[
  \frac {g_{e \rho} g_{f \lambda} (s-M_Z^2)}
        {(s-M_Z^2)^2 + \Gamma_Z^2 M_Z^2}
+ \frac{Q_e Q_f} s
\right]
  \times
\begin{cases}
7(1),     & \rho=\lambda,  \\
1(7),     & \rho\ne\lambda.
\end{cases}
\end{align}
\label{eq:xsec:fb}
\end{subequations}
%
%One may see that the differing $s - M^2_Z$ dependence also appears in $\sigma_{\rm F}$ and $\sigma_{\rm B}$. 
The right panel of \gfig{fig:obs-lo-nlo} shows
the FB asymmetry $A_{\rm FB}$ for the charm quark
final state as a function of the collision
energy $\sqrt s$. The NLO effect introduces an overall negative $A_{FB}$ contribution. NP contribution from SM-dim6 interference is zero at the $Z$ pole, and about linear away from the pole. It is interesting to observe that
$A_{FB}$ reduces after including the NLO
radiative correction. Nevertheless,
the FB asymmetry can reach
5\%-11\% within 3\,GeV offset from the $Z$-pole
which offers a sensitive probe to possible NP
contribution.

\section{Uncertainties and Projected Sensitivities}
\label{sec:uncertainty}

As elaborated above, the new physics contributions
are extracted from the difference between the total
asymmetry and the SM contribution for both the cross
section and forward-backward asymmetries. Since the
predicted new physics contribution with $\Lambda = 1$\,TeV
is at percentage level as shown in the right panels
of \gfig{fig:symmetrization} and \gfig{fig:obs-lo-nlo}.
The new physics scale at $\mathcal O($TeV) summarized
in \gtab{tbl:coe} is actually the one already reached 
at LEP. It is reasonable to expect much better
sensitivity at future lepton colliders. Even with
just an order of improvement on the new physics scale,
the SM-NP interference term will reduce by two orders
to $0.01\%$ level. In order to reach such sensitivity,
the uncertainties need to be controlled to comparable
level.

There are various uncertainties including statistical,
experimental, and theoretical uncertainties.
If different uncertainties do not correlate with
each other, their contributions add up as,
\begin{equation}
  \delta A^2
=
  \delta A_{\rm stat}^2
+ \delta A_{\rm exp}^2
+ \delta A_{\rm th}^2,
\end{equation}
with $\delta A_{\rm stat}$, $\delta A_{\rm exp}$, and
$\delta A_{\rm th}$ denoting the statistical,
experimental, and theoretical uncertainties, respectively.
The correlated uncertainties, especially the experimental
and theoretical uncertainties, will be elaborated later.

\subsection{Efficiency and Statistical Uncertainty}

Although the inclusive and forward-backward asymmetries
are theoretically defined in terms of cross sections
in \geqn{eq:xsec:asym:gen} and \geqn{eq:AFB},
experimentally the asymmetries are evaluated in terms of
event numbers $N_\pm$,
$A \equiv (N_+ - N_-) / (N_+ + N_-)$ \cite{Musolf:1993tb}.
Both event rates $N_+$ and $N_-$ are subject to statistical
uncertainties $\Delta N_\pm = \sqrt{N_\pm}$. Consequently,
the asymmetry is also subject to statistical uncertainty, 
\begin{equation}
  \delta A_{\rm stat}
\equiv
  2 \sqrt{\frac{N_+N_-}{\brc{N_++N_-}^3}}, 
\label{eq:dA:stat}
\end{equation}
with uncorrelated statistical uncertainties $\Delta N_\pm$
which applies for both inclusive and forward-backward
asymmetries. For the inclusive asymmetry
$A_\Delta$, $N_\pm \equiv N\brc{\Delta_\pm}$ are the
total event numbers at the two energy points
$\sqrt s = M_Z \pm \Delta$. For the FB asymmetry
$A_\mathrm{FB}(\sqrt s)$, $N_{+,-} \equiv N_{\rm F,B}(\sqrt s)$ are the total forward and backward
event numbers at a single energy point $\sqrt{s}$.

The total event number $N_\pm = L_\pm \sigma_\pm$ is a
product of the cross section $\sigma_\pm$ and the
corresponding integrated luminosity $L_\pm$. As summarized
in \gtab{tbl:cepc:zscan}, CEPC assigns 28\,ab$^{-1}$
for the $Z$-pole run and 1\,ab$^{-1}$ for each
off-$Z$-pole energy points. For comparison, the FCC-ee
can achieve an integrated luminosity of
at least 150\,ab$^{-1}$ within $\sim$3\,GeV around
the $Z$ pole with two interaction points. To be more
specific, the luminosity is distributed 
$L_0=\brc{40,70,40}$\,ab$^{-1}$ at three
energy points $\sqrt{s}=\brc{87.9,91.2,94.3}$\,GeV
\cite{Janot:2015gjr}. For illustration, we take
the CEPC setup in our discussions below. In order
to project the sensitivity and especially the dependence
on the energy offset $\Delta$, we take 1\,ab$^{-1}$
luminosity for two energy points above and below the
$Z$ pole. In other words, while the luminosity is
fixed to 1\,ab$^{-1}$, the symmetric energy offset
$\Delta$ can freely adjust. Our final sensitivity
estimation is obtained after figuring out the optimal
energy offset $\Delta$ as we elaborate below.

The event rate and the resulting statistical uncertainty
strongly depends on the signal efficiency which is
subject to the signal selection, detector coverage,
as well as muon and hadron jet selection efficiency.
For the forward-backward asymmetry which requires
flavor and charge identification, the event rate
substantially depends on the flavor and charge tagging
efficiencies. This is particularly important for the $b$
and $c$ jet while the muon tagging is almost perfect
and hence can be assumed to have 100\% efficiencies.

For the $b$ and $c$ jets, the flavor and charge tagging,
we take the tagging confusion matrix from the Table 4 of
\cite{Liao:2022ufk}. A flavor tagging efficiency of
$80\%$ (90\%) applies for the $c\bar c$ ($b\bar b$) channel,
respectively. The jet charge tagging efficiencies are
39\% (20\%) for the $c \bar c$ ($b\bar b$) events and
can future improve to $45\% (37\%)$ \cite{Cui:2023kqb}
of which we take the latter as our benchmark value.

\subsection{Theoretical Uncertainties}

The theoretical uncertainty stems from two parts:
1) the uncertainties propagated from the input
parameters which are the so-called parametric
uncertainties, and 2) the missing of higher order
corrections.

\subsubsection{Parametric Uncertainties}

The EW section contains three major parameters,
the two gauge coupling constants $e$ and $g$ in addition
to the Higgs vacuum expectation values (VEV). They can
be equivalently replaced by the fine structure constant
$\alpha \equiv e^2/4 \pi$, the Fermi constant $G_F$, and
the $Z$ mass $M_Z$. Of them, the Fermi constant is currently
the most precisely measured with only $10^{-7}$ relative
uncertainty \cite{ParticleDataGroup:2022pth} and hence can
be omitted for simplicity.

For the fine structure constant, the dominant uncertainty comes from the 5-quark flavor hadronic vacuum
polarization $\Delta\alpha^{(5)}_\text{had}$ \cite{ParticleDataGroup:2022pth}
\begin{equation}
  \Delta\alpha^{(5)}_\text{had}\brc{M_Z}
=
  0.02768 \pm 0.00007,
\label{eq:dalh5}
\end{equation}
which propagate its uncertainty to the running
QED coupling \cite{Jadach:1999vf} through
\begin{align}
  \alpha\brc{s}
=
  \frac {\alpha\brc{0}}
        {1-\Delta\alpha\brc{s}-\Delta\alpha^{(5)}_\text{had}\brc{s}},
\end{align}
and contribute correspondingly a relative error of $7\times 10^{-5}$ to the running fine structure constant at the $Z$ pole scale. This further introduces relative uncertainty at the level of $10^{-4}$ to the $A_{FB}$ and $A_{Pol}$ at off-$Z$-peak points. 
The uncertainty of $\Delta\alpha_{had}^{(5)}$ comes partially
from the $e^+e^-$ annihilation measurement below $\sqrt{s}=2$\,GeV
and partially from the missing higher order calculations of
RG running to the $Z$ pole.
\cite{ParticleDataGroup:2022pth}. 
With future input from low energy $e^+e^-$ collision data, the uncertainty is expected to be improved to around $5\times 10^{-5}$ or better \cite{Janot:2015gjr}.
In future,
a direct measurement of $\alpha(M_Z)$ becomes possible for
the first time at the proposed FCC-ee off-$Z$ run with the order of $\mathcal{O}$(100 ab$^{-1}$) integrated luminosity.
Measuring the forward-backward asymmetry $A_{FB}^{\mu\mu}$
at the two off-$Z$ points is sensitive to the $Z/\gamma$
interference contribution from which $\alpha(M_Z)$ can be
extracted with a relative uncertainty of $3\times 10^{-5}$
\cite{Bernardi:2022hny}. Estimated with ZFitter \cite{Arbuzov:2005ma},
the $\alpha(M_Z)$ error contributes a $10^{-4}$ exact uncertainty
to the off-$Z$-pole polarization and FB asymmetry
$A^{\mu,j}_{\rm pol,FB}$, which serves as the dominant source.

The current and future precisions of the $Z$ boson mass $M_Z$
have been summarized in \gtab{tbl:systematics} for comparison.
In addition, the $Z$ decay width $\Gamma_Z$ also enters the
scattering cross sections as shown in \geqn{eq:xsecs}
and \geqn{eq:xsec:fb}. In principle, it is necessary to do a
joint fit with both EW parameters and the effective operators
for the $Z$ lineshape. For simplicity, we treat the EW parameters
as already determined from the lineshape scan and concentrate
on estimating the consequence of their uncertainties on the
new physics search. The $M_Z$ uncertainty contributes to
$A^{\mu,j}_{\sigma}$ at the level of $10^{-4}$, which
can be reduce by about 20 times with the future $Z$-pole scan data, yet remain the dominant source of parametric uncertainty for the observable.

Different from the $Z$ boson mass $M_Z$ that is used as an
input parameter, the decay width $\Gamma_Z$ is a derived variable.
Instead, the strong interaction coupling strength $\alpha_s(M_Z)$
at the $M_Z$ scale can enter $\Gamma_Z$ through the $Z$ hadronic
decay modes. The current PDG value
$\alpha_s(M_Z) = 0.1180\pm 0.0009$ is expected to reduce to 
$0.00015$ with a precisely measured ratio of the leptonic and
hodronic branching ratio $R_\ell$ \cite{FCC:2018evy}.
With a relative error at almost the $10^{-3}$ level,
$\delta \alpha_s(M_Z)$ dominates the uncertainty for
the cross section asymmetry $A^{\mu,j}_{\sigma}$. From
the current to the future expectation, the corresponding
uncertainty reduces from $10^{-5}$ to $10^{-6}$.

\begin{table}[t]
  \centering
 \begin{tabular}{l|cccc}
    &CEPC & FCC-ee & ILC & Current \\
    \hline
    $\delta M_Z$ (MeV) & 0.1 & 0.1 & 0.2  & 2.1 \\
%    $\delta\Gamma_Z$ (MeV) (\qzn{derived}) & 0.025 & 0.1 & & 2.3 \\
    $\delta\alpha_s(M_Z) $ & $0.00015$ & $0.00015$ & $0.0005$ &  0.0009 \\
    $\delta M_t$ (MeV) & 25\cite{lizhan2023} & 17 & 16 & $\sim 300$ \\
    $\delta \alpha/\alpha(M_Z)$ \cite{Janot:2015gjr} & $5\times 10^{-5}$ & $3\times 10^{-5}$ & $5\times 10^{-5}$ & $1.1\times 10^{-4}$\cite{Bernardi:2022hny} \\
    $\delta M_h$ (MeV) & 5.9\cite{An:2018dwb} & 4.3 & 14 & $\sim 170$ \\ 
    $\delta L/L$  & $5\times 10^{-5}$ & $5\times 10^{-5}$ & $ 10^{-4}$ & $3.4\times 10^{-4}$ \\
    $\delta \sqrt s$ (MeV) & 0.1 & 0.1 & 0.2 & 1.7 \\
  \end{tabular}
\caption{The projected SM input parameter and experimental configuration uncertainties at future colliders (CEPC \cite{CEPCPhysicsStudyGroup:2022uwl},
FCC-ee \cite{Bernardi:2022hny}, ILC\cite{Horiguchi:2013wra,Schwienhorst:2022yqu,Belloni:2022due}) and the existing results
from PDG2022 \cite{ParticleDataGroup:2022pth}
for comparison. The theoretical parametric uncertainties include those for 
 the $Z$ mass ($\delta M_Z$)
and decay width ($\delta \Gamma_Z$), the
Higgs boson mass ($\delta M_h$), and the top quark mass
($\delta M_t$). The experimental
systematic uncertainty for the luminosity ($\delta L$)
and collision energy $\delta \sqrt s$ are shown
together for completeness.
}
\label{tbl:systematics}
 \end{table}

Further, the Higgs mass $M_h$ and top mass $M_t$ can
also enter through loop corrections  \cite{TwoFermionWorkingGroup:2000nks}. 
Especially, their uncertainties \cite{ParticleDataGroup:2022pth},
\begin{equation}
  M_h
=
  125.25 \pm 0.17\,\GeV,
\qquad
  M_t
=
  172.69 \pm 0.3\,\mbox{GeV},
\end{equation}
have sizeable contribution among parametric
uncertainties \cite{TwoFermionWorkingGroup:2000nks}.
The current uncertainty contributions from $M_h$ and $M_t$
are smaller or about the level of $10^{-6}$ and $10^{-5}$
across the three asymmetry observables.
They can significantly reduce with the future lepton collider
Higgs-struglung data and top threshold scan data by about
40 and 20 times, respectively, with uncertainty contributions
of at most $10^{-7}$ and $10^{-6}$ to be expected.

Overall, a $\chi^2$ summation of all dominant parametric
errors discussed above ($\delta M_t$, $\delta M_h$,
$\delta\alpha(M_Z)$, $\delta M_Z$, and $\alpha_s(M_Z)$)
give an overall uncertainty of $\mathcal{O}(10^{-6})$
for all the three off-$Z$-pole asymmetry observables.
Comparing with the theoretical systematic uncerntainties
from the missing higher-order calculations detailed below,
the parametric uncertainties would not be a dominant issue.

\subsubsection{Theoretical Systematics from Missing Higer-Order Calculations}

In the framework of perturbative calculation
with Feynman diagrams, a concrete theoretical
calculation has to truncate at some order.
With missing higher-order calculations, the
difference between the true value of an observable
and the truncated theoretical calculation would
introduce some theoretical systematics. One may
estimate their values according to the existing
calculations at lower orders based on the observation
that higher-order corrections are typically smaller
than the corresponding lower-order counterparts.
As a conservative estimation, one may simply take
the relative size of the existing highest order
calculation as the theoretical systematics from the
missing higher-order corrections.

The state-of-art calculation on the radiative correction
to the $Z$-pole observable is available partially up to the
order \{$\alpha^3, \alpha^2 \alpha_s$\}. For the the weak mixing
angle $\sin^2\theta_W^\text{eff}$ at the $Z$ pole, the theory uncertainty associated with missing higher orders is estimated to be around $4.5 \times 10^{-5}$ \cite{Dubovyk:2018rlg, Chen:2021krh}, corresponding to a relative uncertainty of $\sim 2\times 10^{-4}$.
The weak mixing angle uncertainty directly affects the
forward-backward asymmetry $A_{\rm FB}$. 

When moving off the $Z$ pole, further input and improvement
are needed for estimating the uncertainty of $A_{\sigma, \rm pol}$
\cite{Dubovyk:2019szj,deBlas:2024bmz}as well as the other asymmetries considered in the present work. 
To our knowledge, there exists no complete calculation of NNLO electroweak radiative corrections (EWRC) for the $e^+e^-\to f{\bar f}$ observables of interest here, so we must infer the magnitude of this uncertainty from other work. The aforementioned estimate on the theory error on $\sin^2\theta_W^\text{eff}$ provides one guide. In the case of low-energy, parity-violting (PV) M\o ller scattering, a complete computation of the NNLO  EWRC arising from closed fermion loops has been reported in Ref.~\cite{Du:2019evk}. The result indicates these corrections introduce a $\sim 5\%$ shift in the value of weak charge of the electron in comparison to the NLO result. For $e^+e^-\to \mu^+\mu^-$ off the $Z$-pole, NLO electroweak box contributions induce a $\sim 0.5\%$ correction to the cross section at $\sqrt{s}=189$ GeV\cite{TwoFermionWorkingGroup:2000nks}. Should the ratio of NNLO to NLO corrections follow the same pattern as in low-energy PV M\o ller scattering, one might expect the magnitude of NNLO corrections to $\sigma(e^+e^-\to\mu^+\mu^-)$ to enter at the few $\times 10^{-4} $ level. 
We thus estimate the
missing higher-order theory uncertainties for the three asymmetry
observables assuming uncorrelated relative uncertainty of $10^{-4}$ for the plus and minus cross sections. 
Our estimate could be optimistic, and the impact of missing higher orders could turn out to be larger. As we show below, a $10^{-4}$ relative uncertainty already introduces a significant impediment to realizing the NP mass reach that would be achievable based on the expected experimental, statistical precision.
Given the importance of the off $Z$-pole NNLO EWRC, we hope that their complete computation wil become available in the near future.

We also note that sizeable uncertainties enter 
the $e^+e^-\to q\bar q$ process, most significantly from the
modeling of the jet fragmentation, the final-state QCD radiation,
and the jet-flavor tagging efficiency. For example, the current
estimation of such systematic uncertainty contribution to
$A_{\rm FB}^b$ is about $9\times 10^{-4}$ at the $Z$ pole
\cite{guerreri2023}. Uncertainty at the same level can also
apply to $A_\sigma$ and $A_{\rm pol}$ in the $q\bar q$ channel. One may expect updated theory calculation and simulation
to keep such uncertainty contributions under control. To be
more exact, it should reach at least the $10^{-4}$ level that
is comparable to the missing higher-order calculations at the
parton level.

More discussions and comparison of the experimental,
theoretical input
parameter and missing higher-order uncertainties on a
few important electroweak precision observables at the future
$Z$-factory can be found in the Table 1 of
Ref.\cite{Freitas:2021oiq}. Our analysis echo the conclusion
there that the dominant contribution arises from the missing
higher-order calculation. In addition, \gtab{tab:err_corr}
summarizes the correlation pattern of theoretical uncertainties on
the asymmetry observables. While the statistical uncertainties
affect the two event numbers $N_\pm$ independently and hence
are uncorrelated, the parametric and missing higher order
uncertainties belong to the correlated category by entering
$N_\pm$ simultaneously.

\subsection{Experimental Systematics and Correlations}

There are various sources of experimental systematics,
including luminosity $\delta L$ and the collision
energy resolution $\delta \sqrt s$. With multiple
energy points scan around the $Z$ pole as summarized
in \geqn{tbl:cepc:zscan}, both luminosity and
collision energy have multiple systematics.
\begin{table}[t]
    \centering
    \begin{tabular}{c|ccc|ccc}
Correlation Pattern & $\delta N_\pm$ (stat) &  $\delta N_\pm$(para) &  $\delta N_\pm$(higher) & $\delta N_\pm (\delta L)$ & $\delta N_\pm(\delta \sqrt s)$  &  $\delta N_\pm$ (Pol) \\
\hline
    $A_\sigma$  & $\times$ &  $\checkmark$ & $\checkmark$ & $\times$ & $\times$ &   \\
    $A_{\rm FB}$   & $\times$ &  $\checkmark$ & $\checkmark$ &  & $\checkmark$ &   \\
    $A_{\rm pol}$   & $\times$ &  $\checkmark$ & $\checkmark$ & $\times$ & $\times$ & $\times$
    \end{tabular}
\caption{The effect of the statistical (stat), parametric (para), theoretical systematics from
missing higher-order calculations (higher), and
experimental systematic uncertainties ($\delta L$
for luminosity and $\delta \sqrt s$ for collision
energy) on the
asymmetry observables $A_\sigma$, $A_{\rm FB}$,
and $A_{\rm pol}$ with ($\checkmark$) or
without ($\times$) correlation between the two
event rates $N_+$ and $N_-$ used to define the
asymmetry observables.}
\label{tab:err_corr}
\end{table}
\gtab{tab:err_corr} summarizes the uncertainties
and their effects on the observations. In addition
to the inclusive asymmetry $A_\sigma$ and the
FB asymmetry $A_{\rm FB}$, we also include the
polarization asymmetry $A_{\rm pol}$ elaborated in
\gsec{sec:polarization} for completeness and easy
comparison. The correlation patterns related to
$A_{\rm pol}$ will also be discussed there.

On the experimental side, the luminosity uncertainty
can have both correlated and uncorrelated parts.
However, the correlated one affects both
$N_+ = N(M_Z + \Delta)$ and $N_- = N(M_Z - \Delta)$
and hence cancels out in the inclusive asymmetry
$A_\sigma$ which is defined as a ratio of event
rates $N_\pm$. So $A_\sigma$ is affected by only
the uncorrelated luminosity uncertainty. For the
forward-backward asymmetry $A_{\rm FB}$ that is
defined at a single energy point, the luminosity
uncertainties all cancel out and there is no need
to put a correlation pattern for this item in
\gtab{tab:err_corr}. For the collision energy,
each point has its own uncertainty $\delta \sqrt s$.
Being defined at two energy points, the inclusive
asymmetry $A_\sigma$ is subject to uncorrelated
$\delta \sqrt s$. On the contrary, the FB asymmetry
$A_{\rm FB}$ is defined at single collision energy
and hence the single $\delta \sqrt s$ affects both
event rates $N_\pm$ with full correlation.

The experimental precision goals at the CEPC
and FCC-ee in comparison with the LEP precision
are summarized in \gtab{tbl:systematics}. The
luminosity uncertainty reduces from the LEP
achievement by at least 3 times to 0.01\%, and the uncorrelated luminosity between energy points is further expected to reach below $5\times 10^{-5}$
\cite{Bernardi:2022hny}.
Since only the uncorrelated luminosity uncertainty
can leave effect in asymmetry observations, its
value shown is for the uncorrelated one.
For comparison, the collision energy uncertainty
$\delta \sqrt s$ improves by more than 10 times
from LEP to only 0.1\,MeV \cite{Blondel:2019jmp,Bernardi:2022hny}.

According to the error propagation rule, the
uncorrelated uncertainties affect the asymmetry observables ($A_\sigma$, $A_{\rm FB}$, $A_{\rm pol}$)
in the following way,
\begin{equation}
  \delta A^2
\equiv
  \frac {4 N_-^2 (\delta N_+)^2}
        {(N_+ + N_-)^4}
+
  \frac {4 N_+^2 (\delta N_-)^2}
        {(N_+ + N_-)^4},
\qquad
  \delta N_\pm
\equiv
  \frac{\partial N_\pm}{\partial X_\pm} {\delta X_\pm},
\label{eq:uncorr}
\end{equation}
which $\delta X_\pm = \left\{\delta\Delta,~\delta L, \delta {\rm Pol} \right\}$.
The error propagation for the correlated uncertainties,
such as the energy resolution on $A_{\rm FB}$, is much
simplier,
\begin{equation}
  \delta A
=
  \frac{\partial A}{\partial X}
  \delta X
=
  \frac {2 \, \delta X} {(N_+ + N_-)^2}
\left(
  N_- \frac {\partial N_+}{\partial X}
- N_+ \frac {\partial N_-}{\partial X}
\right),
\end{equation}
Uncorrelated means there are two uncertainties
$\delta X_\pm$ of the same type but without
correlation while correlated means the two
uncertainties $\delta X_\pm$ are actually the
same $\delta X$.

\begin{figure}[t]
\centering
\includegraphics[width=0.48\textwidth]{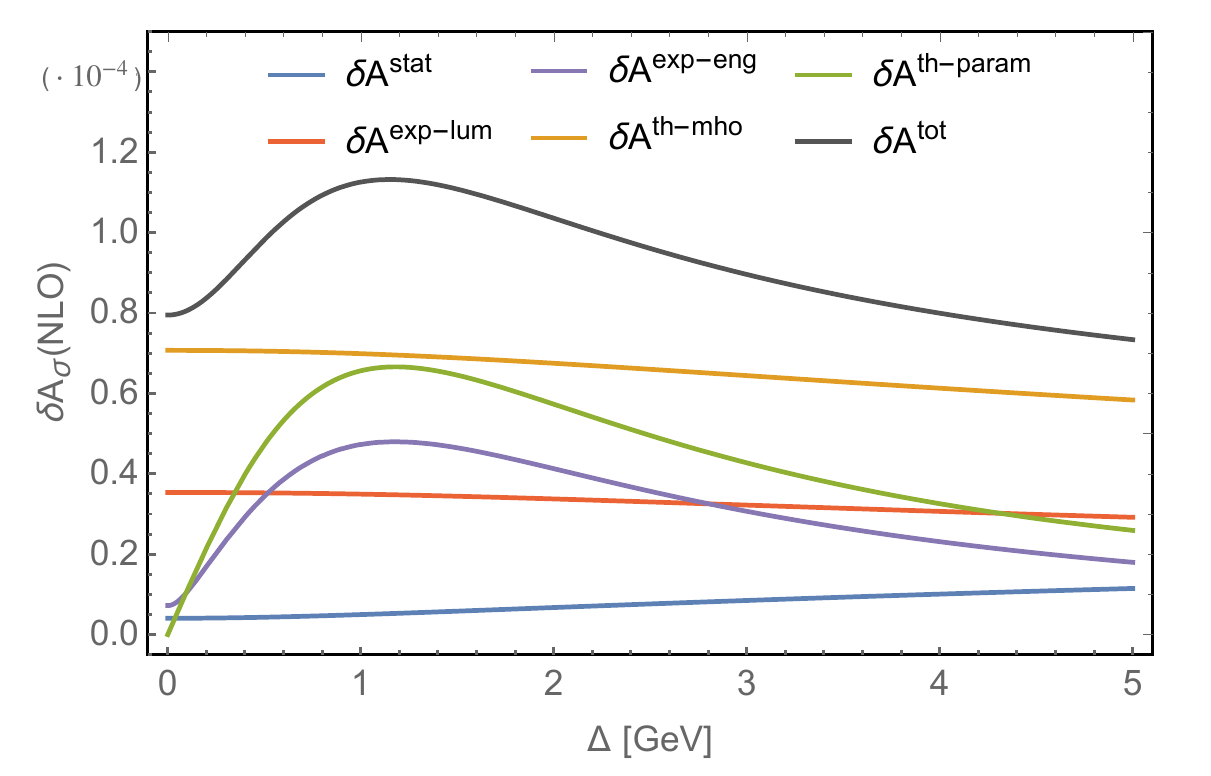}
\includegraphics[width=0.48\textwidth]{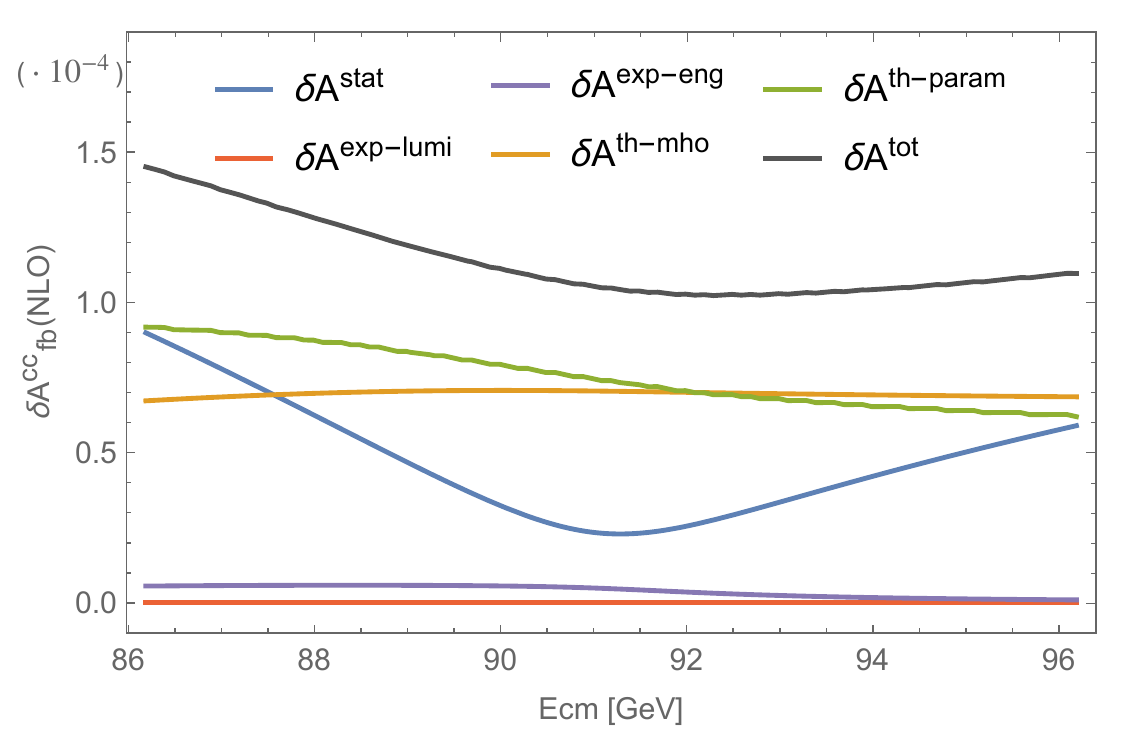}
\caption{The effect of various uncertainties
(statistical $\delta A^{\rm stat}$, the experimental
luminosity $\delta A^{\rm lum}$ and collision energy
$\delta A^{\rm eng}$, as well as the theoretical
$\delta A^{\rm th}$) on the inclusive asymmetry
$\delta A_\sigma$ (Left) and FB asymmetry
$\delta A_{\rm FB}$ (Right). For comparision, the
combined experimental uncertainty $\delta A^{\rm exp}$
and the total uncertainty $\delta A^{\rm tot}$
are also shown. These results take the CEPC
luminosity projection as a conservative illustration.}
\label{fig:deltaA}
\end{figure}
The effect of various uncertainties on the
inclusive and FB asymmetries are illustrated
in \gfig{fig:deltaA}. With significantly enhanced
event rate at future lepton colliders, the statistical
uncertainty is the least important. The experimental
uncertainties, including both luminosity and collision
energy systematics, dominates at small offset $\Delta$
for the inclusive asymmetry. With increasing offset,
the luminosity uncertainties persists to be almost
constant while the collision energy counterpart
descreases beyond $\Delta \approx 1$\,GeV. However,
the experimental uncertainties have almost negligible
effect, which is even smaller than the statistical
one, for the FB asymmetry across the whole energy range.
Comparing with the high statistics and significantly
improved experimental uncertainties, the theoretical
uncertainties contribute the largest share with
domination in the range of $\Delta \gtrsim 1.5$\,GeV
for the inclusive asymmetry $A_\sigma$ and cross the
whole range for the FB asymmetry $A_{\rm FB}$. It is
of crucial importance to reduce the theoretical
uncertainties to match the experimental precision
at future lepton colliders to fully benefit from their
potentials. If achieved, both the inclusive $A_\sigma$
and the FB $A_{\rm FB}$ asymmetries are guaranteed to
reach $10^{-5}$ precision.

\subsection{Projected Sensitivities at CEPC}

\begin{figure}
\centering
\includegraphics[width=0.48\textwidth]{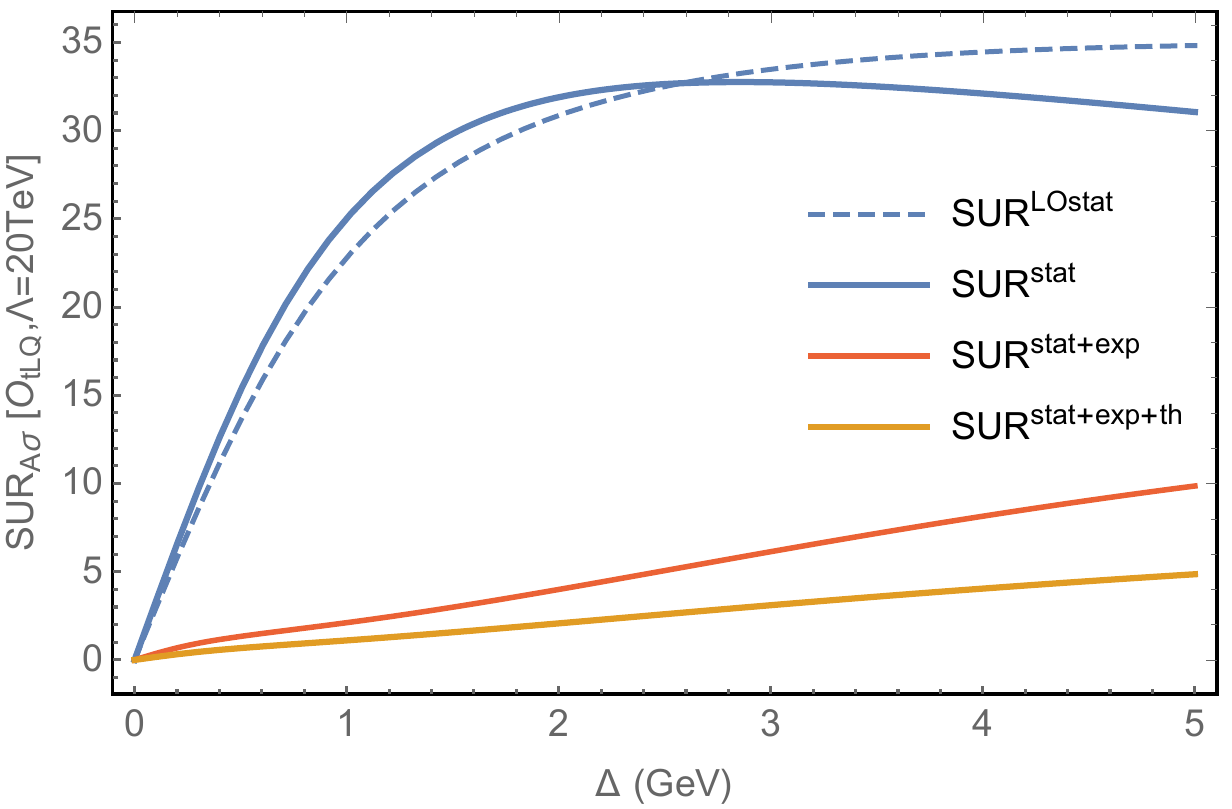}
\hfill
\includegraphics[width=0.48\textwidth]{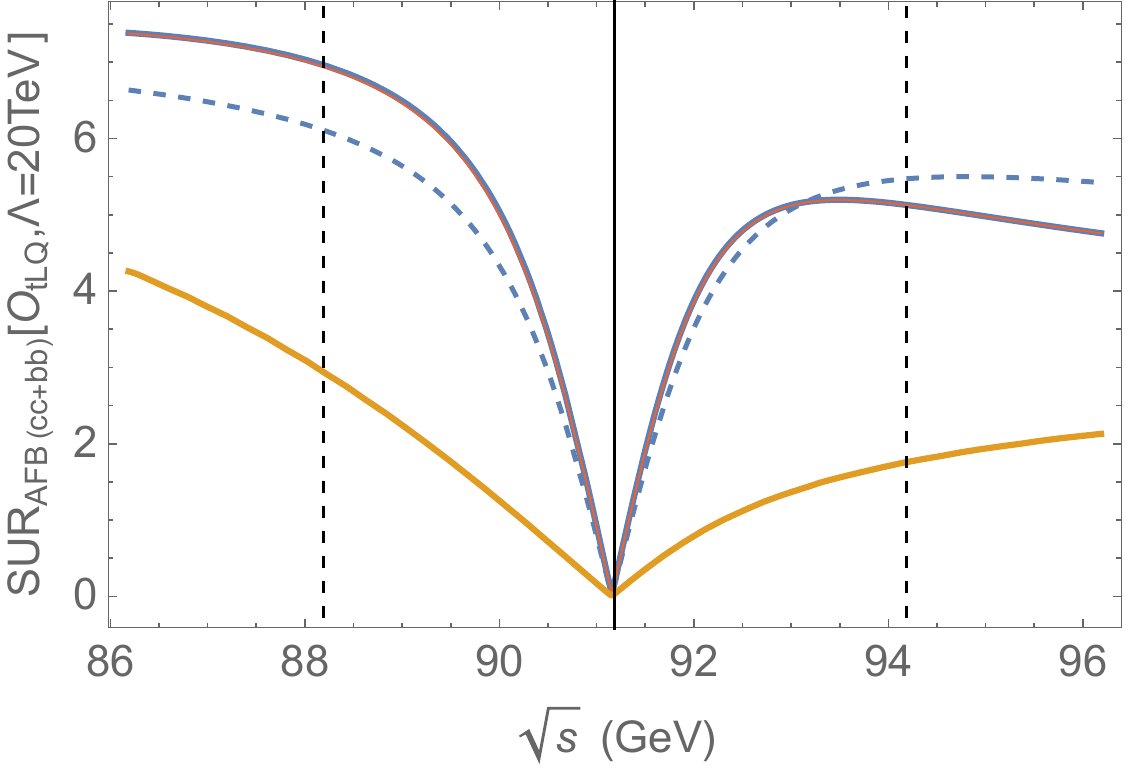}
\caption{The signal-uncertainty-ratio (SUR) for the
$O_{LQ}^t$ operator with $\Lambda=20$\,TeV
projected for the inclusive $A_\sigma$ (Left)
and FB $A_{\rm FB}(cc+bb)$ (Right) asymmetries at CEPC.}
\label{fig:sig12}
\end{figure}

The projected signal-uncertainty-ratio (SUR) at CEPC can be evaluated
by comparing the signal strength as elaborated in
\gsec{sec:observable} with the uncertainties
discussed in the previous subsections. For illustration,
we take $\mathcal{O}^s_{LQ}$ operator with
$\Lambda = 20$\,TeV as benchmark. Such operator can
have quark final states in the
$e^+ e^- \rightarrow q \bar q$ scattering.
The expected
SUR $n(\sigma) \equiv A_i^{\rm NP}/\delta A_i$
is then a function of the collision energy offset
$\Delta$ from the $Z$ pole as shown in \gfig{fig:sig12}.
Although the patterns shown in \gfig{fig:sig12} and
elaborated in the following two paragraphs are
obtained with illustration of the $\mathcal{O}^s_{LQ}$
operator, similar patterns also apply for the other
10 operators listed in \gtab{tbl:11opnew}.

To demonstrate the effect of various uncertainties,
we show the signal SUR with different
uncertainty combinations. With only statistical
uncertainties, the two curves with LO (dashed blue)
and NLO (solid blue) theory prediction are quite
close to each other.
Comparing with \gfig{fig:obs-lo-nlo}, the inclusion
of NLO calculation mainly affects the SM prediction
but not the NP SUR. Including the experimental
systematics can significantly affect the SUR
from the inclusive asymmetry $A_\sigma$
measurement but the one from the FB asymmetry
$A_{\rm FB}$ is not affected.
Finally, the theoretical uncertainties can further
reduce the SUR which is most
significant for $A_{\rm FB}$ across the whole range
and the large offset region for $A_\sigma$.
These observations are consistent with the
uncertainty patterns in \gfig{fig:deltaA}.

For the collision energy offset $\Delta$ dependence,
the projected SUR increases with $\Delta$ if all
known uncertainties including the current theoretical
uncertainties are taken into consideration. However,
the theoretical uncertainties have the prospect of
possible improvement to even better than the
experimental systematics for the inclusive asymmetry
$A_\sigma$ and probably the statistical uncertainty
for the FB asymmetry $A_{\rm FB}$. Then more features
need attension. For $A_\sigma$, the SUR first
increases fast with the collision energy offset
$\Delta$ for $\Delta \lesssim 2$\,GeV and then
becomes almost flat for larger $\Delta$. On the other
hand, the FB asymmetry SUR also first increases fast
with $\Delta$ for $\Delta \lesssim 1$\,GeV. Rather
than being flat, the $A_{\rm FB}$ SUR keeps increasing
for the collision energy below the $Z$ pole while
the one above starts to decrease. With all these
features taken into consideration, a balanced choice
of the collision energy offset is $\Delta = 3$\,GeV
on both sides of the $Z$ pole which is more or less
consistent with the official scheme for the $Z$
lineshape scan as summarized in \gtab{tbl:cepc:zscan}.

\begin{figure}
\centering
\includegraphics[width=0.48\linewidth]{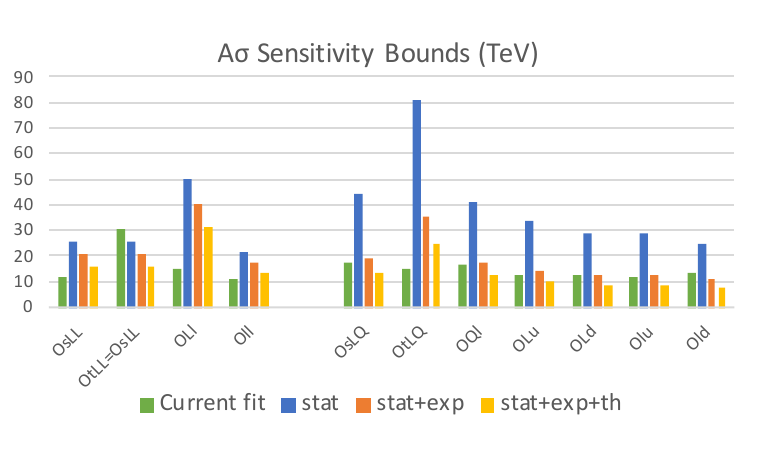}
\hfill
\includegraphics[width=0.48\linewidth]{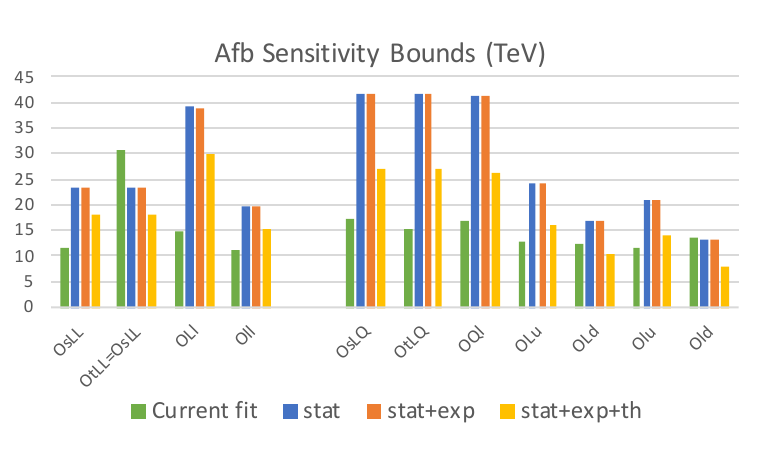}
\caption{The projected $2\sigma$ sensitivity
on the new physics scale from the inclusive
asymmetry $A_{\sigma}$ (Left) and FB asymmetry
$A_{\rm FB}$ (Right) measurements at CEPC. For illustration,
the collision energy offset takes the balanced
choice $\Delta=\pm 3$\,GeV.}
\label{fig:sensitivity}
\end{figure}

The SUR in \gfig{fig:sig12} is obtained with
fixed NP scale $\Lambda = 20$\,TeV for a single
operator $\mathcal O^s_{LQ}$. It is more intuitive
to also show the projected sensitivity for
all operators as done in \gfig{fig:sensitivity}.
The FB asymmetry sensitivity has two contributions
above and below the $Z$ pole which are combined
in the right panel. Both panels show the projected
sensitivities at $2\sigma$ confidence level.

If only statistical uncertainty, the projected
$2 \sigma$ sensitivity can reach $\mathcal O(10)$\,TeV
for all operators. Especially, the $\mathcal O_{L\ell}$
and $\mathcal O^t_{LQ}$ operators can reach
50\,TeV and 80\,TeV, respectively.
The $\mathcal O_{L\ell}$ operator can contribute to
both $e^+_L e^-_L$ (with $e$ from $L$) and
$e^+_R e^+_R$ (with $e$ from $\ell$) beam chirality
configurations. For $\mathcal O^t_{LQ}$, the enhancement
comes from the fact that the cross sections of up-
and down-type final state quarks have the same sign
and add up to give large SUR ratio.

Including the experimental systematics will significant
reduce the $A_\sigma$ sensitivities. Especially,
the 80\,TeV for $\mathcal O^t_{LQ}$ significantly
shrinks to below 30\,TeV. Typically, the leptonic
$\mu^+ \mu^-$ states are less affected than their
hadronic counterparts. These explain why the sensitivity
significantly decreases for the last 7 but only
minor reduction for the first 4 operators in the
left panel of \gfig{fig:sensitivity}
For comparison, the FB asymmetry
$A_{\rm FB}$ is not that sensitive to the
experimental uncertainties as shown in the
right panel of \gfig{fig:sensitivity}. For comparison,
the theoretical uncertainties has the opposite
effects on $A_\sigma$ and $A_{\rm FB}$. Now
$A_\sigma$ is only moderately affected by the
theoretical uncertainties  while the sensitivity
from $A_{\rm FB}$ is significantly reduced.

Between the inclusive asymmetry $A_\sigma$ in
the left panel and the FB asymmetry $A_{\rm FB}$
in the right panel, $A_\sigma$ gives better
sensitivity on almost all the operators with
the only exceptions of $\mathcal O^t_{LQ}$
and $\mathcal O_{Q \ell}$ considering all
uncertainties. Although the FB asymmetry $A_{\rm FB}$
is more frequently studied, its inclusive
counterpart actually has better sensitivities
for the four-fermion NP operators around the
$Z$ pole.

\section{Polarization Asymmetry}
\label{sec:polarization}

Although polarization is in general difficult at circular
lepton colliders, it is potentially feasible around the $Z$ pole
according to various on-going discussions for CEPC \cite{Duan:2023cgu,CEPCStudyGroup:2023quu}
and FCC-ee \cite{Bauche:2023dlq,FCC:2018evy}. Available techniques can hope to deliver
$(f_{e^-}, f_{e^+})$ = (80\%, 65\%) effective
electron left-handed ($f_{e^-}$) and positron right-handed
($f_{e^+}$) beam
polarization fractions
\footnote{The polarization fraction $f_{e^\pm}$ is converted
from the usual definition as
$f_{e^\pm} \equiv (1 \pm P_{e^\pm}) / 2$. A beam polarization of over $20$\% for positron is possible with Sokolov-Ternov effect, and polarization loss in the acceleration can be maintained small at the $Z$-pole center of mass energy\cite{Duan:2023cgu}.}.

With beam polarization, it is possible to define two
different event rates $N_{+-}$ and $N_{+-}$ with 100\%
$e^-_L e^+_R$ and $e^-_R e^+_L$ pure beam polarization
configurations, respectively. With vector interactions
in the SM electroweak currents and those four-fermion operators
defined in \gtab{tbl:11opnew}, the other two combinations
$e^-_L e^+_L$ and $e^-_R e^+_R$ have only vanishing
cross sections and hence can be omitted. In reality,
the beam polarization cannot be pure and is always a hybrid,
\begin{subequations}
\begin{align}
  N_+
& \equiv
  f_{e^-} f_{e^+} N_{+-} 
+ (1- f_{e^-}) (1 - f_{e^+}) N_{-+},
\\
  N_-
& \equiv
  (1 - f_{e^-}) (1 - f_{e^+}) N_{+-} 
+ f_{e^-} f_{e^+} N_{-+}.
\end{align}
\end{subequations}
Between $N_+$ and $N_-$, both the electron and positron
beam polarizations are reversed,
$f_{e^\pm} \leftrightarrow 1 - f_{e^\pm}$. It is then
possible to define a polarization asymmetry $A_{\rm pol}$
between these two beam polarization configurations,
\begin{equation}
  A_{\rm pol}
\equiv
  \frac {N_+ - N_-}{N_+ + N_-}
=
  P_{\rm eff}
  \frac {N_{+-} - N_{-+}}
        {N_{+-} + N_{-+}},
\quad \mbox{with} \quad
  P_{\rm eff}
\equiv
  \frac {f_{e^-}+f_{e^+}-1}
        {1 - f_{e^-} -f_{e^+} + 2 f_{e^-} f_{e^+}}.
\end{equation}
It is interesting to see that the beam polarizations
factorize out as an overall prefactor $P_{\rm eff}$.

\begin{figure}[t]
\centering
\includegraphics[width=0.46\linewidth]{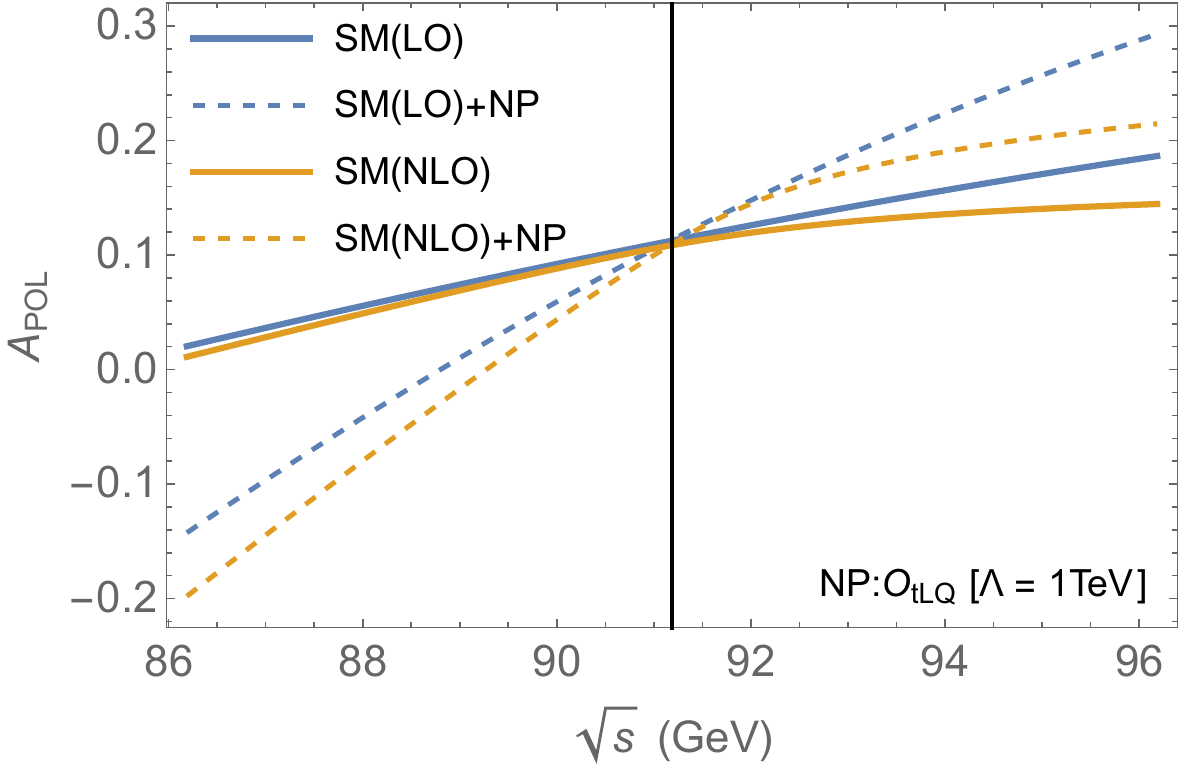}
\hfill
\includegraphics[width=0.49\linewidth]{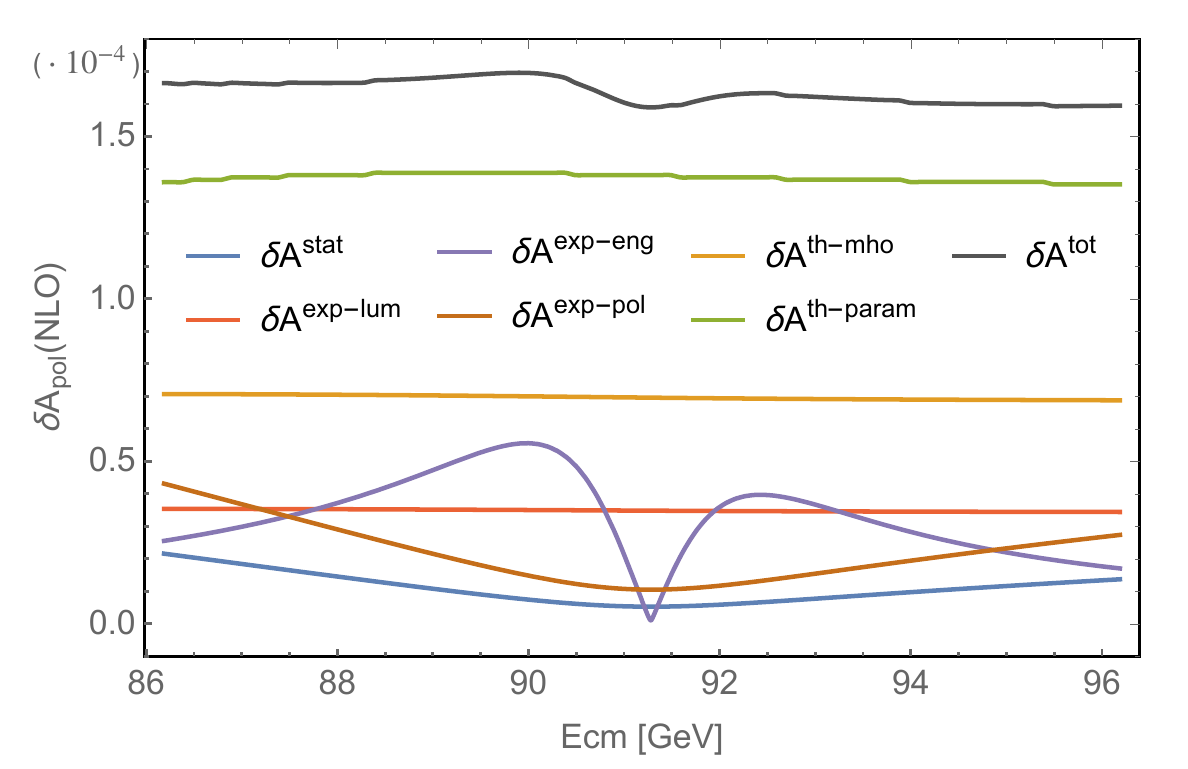}
\caption{The polarization asymmetry $A_{\rm pol}$ (Left)
and its uncertainties $\delta A_{\rm pol}$ (Right)
for the $\mathcal O^t_{LQ}$ operator with $\Lambda = 1$\,TeV. 
For $A_{\rm pol}$, both SM (solid) and SM+NP (dashed)
contributions are shown with LO (blue) and NLO (yellow)
calculations, respectively, for comparison. On the other hand,
the $\delta A_{\rm pol}$ panels illustrates the size
of statistical (stat), experimental luminosity (lumi)
and collision energy (eng) systematic, and
theoretical (th) uncertainties in addition to the total
(tot) one with all uncertainties combined.}
\label{fig:pol12}
\end{figure}

\gfig{fig:pol12} takes the $\mathcal O^t_{LQ}$ operator to
illustrate the polarization asymmetry $A_{\rm pol}$
and its uncertainties $\delta A_{\rm pol}$ in the
left and right panels, respectively. The left panel shows
that its value is at the level of $0\% \sim 20\%$ level
around the $Z$ pole and can be further enhanced to the
range of $-20\% \sim 30\%$ by $\mathcal O^t_{LQ}$ with
$\Lambda = 1$\,TeV in the similar way as the inclusive
$A_\sigma$ and FB asymmetries as shown in \gfig{fig:obs-lo-nlo}. 
One may expect the polarization asymmetry to have
comparable sensitivity on the NP four-fermion operators.

With beam polarization, the luminosity for the $Z$ line
shape scan at each collision energy as summarized in
\gtab{tbl:cepc:zscan} needs to be redistributed for the
two polarization configurations. A reasonable scheme is
dividing the luminosity equally among the two configurations,
each with $L_0 =0.5\,{\rm ab}^{-1}$. Although the event
rates are reduced by half at each point, the statistical
uncertainty is still negligibly small as shown in the
right panel of \gfig{fig:pol12}. With the polarization
asymmetry $A_{\rm pol}$ defined at a single energy point,
the two beam polarization configurations however can not
run simultaneously. Consequently, the luminosity and
collision energy uncertainties are uncorrelated and have
two independent copies for the two polarization configurations.
In other words, these two experimental uncertainties
for $A_{\rm pol}$ are more or less the same as for the
inclusive asymmetry $A_\sigma$ defined between two
collision energy points rather than the FB asymmetry
$A_{\rm FB}$ at a single energy point.
In addition to the
luminosity and collision energy uncertainties
whose effect is relatively small, the major experimental
systematic uncertainty comes from the beam polarization uncertainties.
Unfortunately, there is no reliable estimation about this
polarization uncertainty. So we simply take
$\delta f_{e^\pm} = 10^{-3}$ as requirement to reach
a reasonable sensitivity. With this conservative
$0.1\%$ requirement, the polarization uncertainty
dominates over the other two experimental systematics.
Fortunately, the polarization systematics could
potentially be controlled to the level comparable to the
statistic uncertainty \cite{Blondel:1987wr}. We thus take
the polarization uncertainty to be twice the statistical
uncertainty in the current study.
In short, the polarization asymmetry can help to
identify the NP operators after the three experimental
uncertainties become comparable with each other.
Another major concern is about the theoretical uncertainties
whose current estimation is even larger than the
polarization one.

\begin{figure}
\centering
\includegraphics[width=0.46\linewidth]{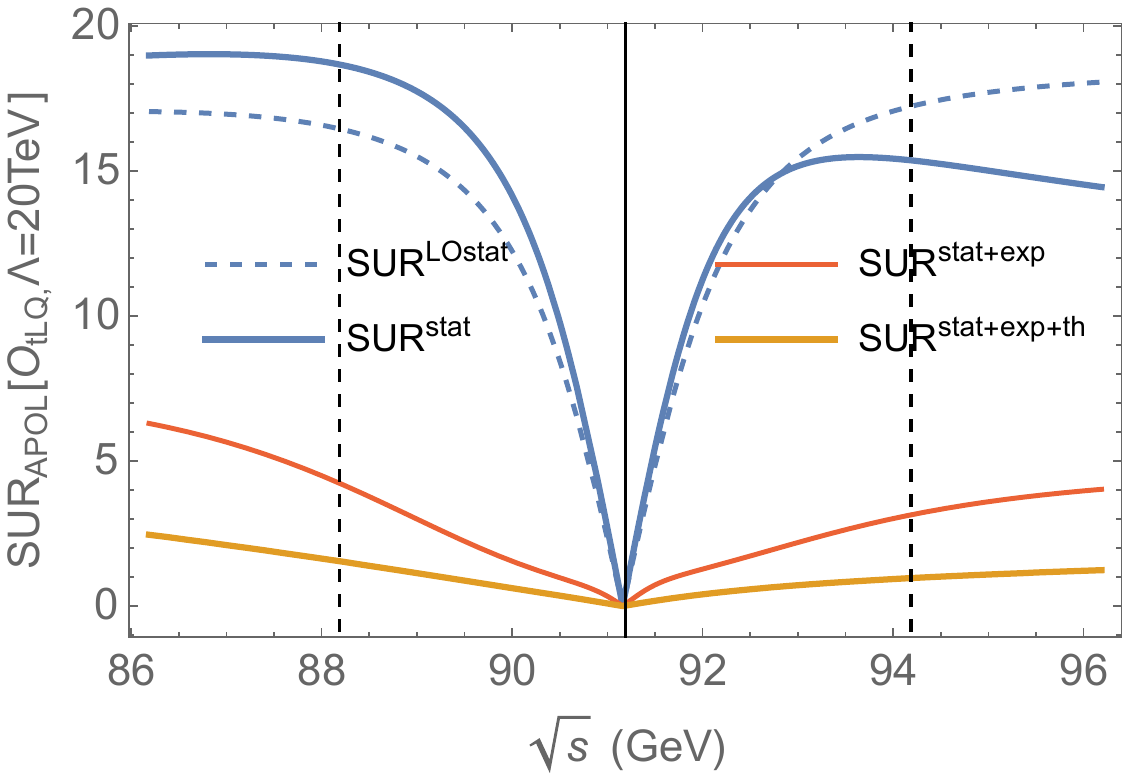}
\includegraphics[width=0.50\linewidth]{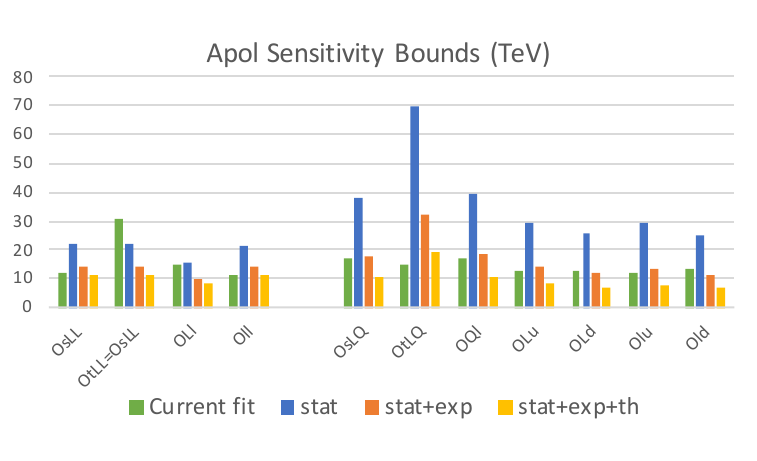}
\caption{(Left) The signal-uncerntainty rato (SUR) as function of the
collision energy $\sqrt s$ for the benchmark $\mathcal O^t_{LQ}$
operator with $\Lambda = 20$\,TeV. (Right) The projected
$2\sigma$ sensitivities on the new physics scale $\Lambda$.
Both panels take the polarization asymmetry observable
$A_{\rm pol}$ (right) with collision energy offsets
$\Delta = 3$\,GeV below and above the $Z$ pole. }
\label{fig:pol34}
\end{figure}

The left panel of \gfig{fig:pol34} shows the projected
SUR with the polarization asymmetry observable $A_{\rm pol}$
at CEPC for the $\mathcal O^t_{LQ}$ operator as an example.
Even with NP scale at $\Lambda = 20$\,TeV, the size of
SUR is quite sizable with only statistical uncertainty
(blue). Although the NLO calculation can significantly
affect the SM contribution as illustrated in the left
panel of \gfig{fig:pol12}, it does not affect the signal
probe. The significant reduction comes from the experimental
and theoretical uncertainties as elaborated right above
for the right panel of \gfig{fig:pol12}. The $\Delta = 3$\,GeV
choice for the energy offset still applies here for the
polarization asymmetry $A_{\rm pol}$. Based on these
observations, the projected sensitivities from $A_{\rm pol}$
are comparable with those for $A_\sigma$ and $A_{\rm FB}$ if there
is only statistical uncertainty but deteriorates
significantly once including the beam polarization
and theoretical uncertainties.

\section{Discussions and Conclusions}
\label{sec:conclusion}

\begin{figure}
\centering
\includegraphics[width=0.68\linewidth]{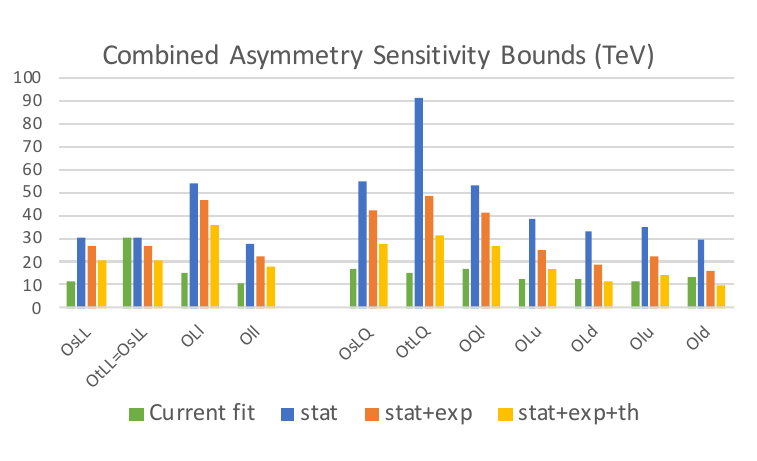}
\caption{The projected $2\sigma$ sensitivities on the new physics
scale $\Lambda$ at CEPC. The inclusive ($A_\sigma$), FB ($A_{\rm FB}$),
and polarization ($A_{\rm pol}$) asymmetry observables with collision
energy offset $\Delta = 3$\,GeV below and above the $Z$ pole are
all combined. Current fit from \cite{Han:2004az} is also included for comparison.}
\label{fig:sensitivitycomb}
\end{figure}

In \gfig{fig:sensitivitycomb} we show the combined $2 \sigma$ sensitivity
projection with three asymmetry observables $A_\sigma$,
$A_{\rm FB}$, and $A_{\rm pol}$. When performing the combination,
we have assumed that the uncertainties are fully uncorrelated
among the three observable. The best sensitivity is associated with
the $\mathcal O^t_{LQ}$ operator which can reach 90\,TeV
if there is only statistical uncertainty. Including experimental and theoretical uncertainties significantly
reduces the sensitivity reach. With the current estimation of
theoretical uncertainties, the best sensitivity shifts to the
$\mathcal O_{L \ell}$ operator reaching 30\,TeV with all
factors considered. Further reducing theoretical uncertainties
may allow the sensitivity to reach 40\,TeV for operators
$\mathcal O_{L \ell}$, $\mathcal O^s_{LQ}$, $\mathcal O^t_{LQ}$,
and $\mathcal O_{Q \ell}$.

With the results obtained in this paper, one may see that
the off $Z$-pole run can play an important role in not just
the $Z$ line shape scan but also the search for new physics
appearing in the guise of four-fermion effective operators. 
The corresponding NP sensitivity arises through the SM-NP
interference that happens only off the $Z$ pole. Our study suggests that the assignment of run time for off $Z$-pole studies would provide a valuable component to the future lepton collider program.
%It is
%beneficial to assign more run time for the off $Z$-pole
%collision energies. 
%In addition, 
%\mrm{Realizing the full physics reach would require that}
%the polarization uncertainty
%\sout{needs to} be controlled below the $0.1\%$ level to allow comparable
%sensitivity reach for the polarization asymmetry $A_{\rm pol}$
%as \mrm{compared to the} inclusive ($A_\sigma$) and FB ($A_{\rm FB}$) counterparts.
Realizing the full physics reach would require the theoretical uncertainties to be
reduced from the current level to match the unprecedented
experimental precision at future lepton colliders. 

\section*{Acknowledgements}

The authors would like to thank Zhe Duan, Zhijun Liang,
Manqi Ruan, and Junping Tian for useful discussions and helps.
The authors are supported by the National Natural Science
Foundation of China grant numbers 12375094 (MJRM), 1240050404 (ZQ) and 11975150 (MJRM).
SFG is supported by the National Natural Science
Foundation of China (12425506, 12375101, 12090060 and 12090064) and the SJTU Double First
Class start-up fund (WF220442604).
JZ and MJRM were also supported under U.S. Department of Energy contract no. DE-SC0011095. SFG is also an affiliate member of Kavli IPMU, University of Tokyo. ZQ would like to thank the host of TDLee Institute of SJTU during her stay as visiting scholar.

\addcontentsline{toc}{section}{References}
%\begin{thebibliography}{99}
%\end{thebibliography}

\bibliographystyle{utphysGe}
\bibliography{refs}
%\nocite{*}

\end{document}